\documentclass[intlimits,twoside,a4paper]{article}

\usepackage{amsmath,amssymb}
\usepackage{graphicx}

\usepackage[T2A]{fontenc}
\usepackage[cp1251]{inputenc}
\usepackage{color}
\usepackage[eqsecnum]{cmpj2}

\issue{2015}{18}{2}{22801}
\doinumber{10.5488/CMP.18.22801}
\articletype{Review article}

\title[Self-assembly of DNA-functionalized colloids]%
{Self-assembly of DNA-functionalized colloids%
}
\author[P.E. Theodorakis \textsl{et al.}]{P.E.~Theodorakis\refaddr{label1}, N.G.~Fytas\refaddr{label2}, G.~Kahl\refaddr{label3,label4}, Ch.~Dellago\refaddr{label4,label5}}

\addresses{
\addr{label1} Department of Chemical Engineering, Imperial College London, South Kensington Campus, \\SW7 2AZ London, UK
\addr{label2} Applied Mathematics Research Centre, Coventry University, CV1 5FB, Coventry, UK
\addr{label3} Institute for Theoretical Physics, Vienna University of Technology, Wiedner Hauptstra{\ss}e 8--10, \\A--1040 Vienna, Austria
\addr{label4} Center for Computational Materials Science (CMS), Wiedner Hauptstra{\ss}e 8--10, A--1040 Vienna, Austria
\addr{label5} Faculty of Physics, University of Vienna, Boltzmanngasse 5, A--1090 Vienna, Austria
}
\date{Received January 21, 2015, in final form February 24, 2015}

\authorcopyright{P.E.~Theodorakis, N.G.~Fytas, G.~Kahl, Ch.~Dellago, 2015}
\begin{document}

\maketitle

\begin{abstract}
Colloidal particles grafted with single-stranded DNA (ssDNA)
chains can self-assemble into a number of different crystalline
structures, where hybridization of the ssDNA chains creates links
between colloids stabilizing their structure. Depending on the
geometry and the size of the particles, the grafting density of
the ssDNA chains, and the length and choice of DNA sequences, a
number of different crystalline structures can be fabricated.
However, understanding how these factors contribute
synergistically to the self-assembly process of DNA-functionalized
nano- or micro-sized particles  remains an intensive field of
research. Moreover, the fabrication of long-range structures due
to kinetic bottlenecks in the self-assembly are additional
challenges. Here, we discuss the most recent advances from theory
and experiment with particular focus put on recent simulation
studies.
\pacs 82.70.Dd, 87.14.gk, 81.16.Dn, 61.46.Df, 61.50.Ah, 75.75.Fk
\end{abstract}

\section{Introduction and overview}

The miniaturization of technology demands new highly efficient
methods to fabricate materials with desired optical, electronic
and mechanical properties. To this end, the self-assembly of nano-
and micron-sized particles functionalized with DNA strands has
attracted much attention as a promising candidate for the
formation of nano-crystals with long-range periodicity
\cite{Crocker2008, Travesset2011,Rogers2015}, as well as in applications such
as sensing, imaging and therapeutics
\cite{Hurst2009,Hurst2011,Zuccheri2011,Heuer2013,Rosi2005,Sokolova2008,Tan2011,Rosi2006,
Niemeyer2001}. The role of DNA strands attached onto the surface
of the particles is to stabilize soft non-compact crystalline
structures by acting as links between the particles or as the
intermediate matter. The structures depend on the size and
geometry of the nano-particles (NPs), the surface density of DNA
strands grafted onto the particles and parameters related to the
DNA strands, such as the number of complementary bases which are
available for hybridization \cite{Macfarlane2014}. Additionally,
the occurrence of kinetically trapped structures during the
self-assembly depends on thermodynamic conditions.

The idea of using DNA strands to create non-compact crystalline
nano-structures was introduced by Alivisatos et al.
\cite{Alivisatos1996} and Mirkin  et al. \cite{Mirkin1996},
where gold (Au) NPs were used. These NPs are also referred to as
polyvalent DNA Au NP conjugates \cite{Hurst2008} due to the large
number of oligonucleotide chains grafted onto the NPs. Since then,
much research is dedicated to understanding the self-assembly
process of DNA-functionalized NPs and designing strategies that
lead to specific crystalline structures. After almost two decades
of intensive research in this field, a very good understanding has
been achieved through a number of most interesting results.
However, significant problems remain unsolved; for example,
kinetic bottlenecks during the self-assembly process may result in
the formation of unwanted crystalline, or disordered structures.
Moreover, synthesis of DNA-coated NPs in a controlled and accurate
manner requires further improvement \cite{Macfarlane2013}.
Therefore, the manufacturing of long-range ordered nano-structures
required for many applications remains challenging. Additionally,
bulk non-compact materials with a photonic band-gap in the range
of visible wavelengths have not been achieved either
\cite{Norris2007}.

Experimental, theoretical and computer simulation methods have
contributed substantially to this area of research. While computer
simulations offer some advantages related to the control of conditions over theory and experiment,
limitations in system size and availability of reliable force
fields underline the importance of a combined approach in
studying the self-assembly of DNA-functionalized
NPs. This review attempts to briefly present the  key results in this
research area with an emphasis put on the simulation side.

Experimental research and a range of applications of
DNA-functionalized NPs have been previously discussed in a number
of
reviews~\cite{Barrow2013,Geerts2010,Heuer2013,Hung2010,Katz2004,Macfarlane2011,Mazid2014,Rosi2005,Storhoff1999,Sokolova2008,Tan2011},
while very little discussion has been reported on the recent
progress of theoretical and simulation contribution to this area
of research \cite{Perez2011,Li2014,DiMichele2013,Zhang2015}. Part
of the experimental discussion has focused on the synthesis,
characterization and phase behavior of DNA-coated NPs assemblies
\cite{Barrow2013,Hung2010,Mazid2014,Storhoff1999,Geerts2010,Katz2004,Tan2011,
Macfarlane2013}, while others have concentrated on the possible
applications of DNA-functionalized colloids
\cite{Heuer2013,Rosi2005,Sokolova2008}.

Storhoff and Mirkin \cite{Storhoff1999} have described the
synthesis of DNA-programmed materials \cite{Cao2001} referring to
meso- and macroscopic organic structures from DNA at its infancy.
Later, Macfarlane  et al.  have synopsized the key findings
of their work suggesting fundamental design rules to predict the
synthesis of distinct colloidal crystal structures with lattice
parameters within the range 25--150~nm based on the independent
adjustment of particle size, lattice periodicity and
inter-particle distance \cite{Macfarlane2011,Macfarlane2010,
Macfarlane2013}. For example, it was suggested
\cite{Macfarlane2011,Macfarlane2010} that the overall hydrodynamic
radius of a DNA NP dictates the assembly and packing behavior of
the system, where the hydrodynamic radius is the size of the NP
with the oligonucleotide components. NPs of equal hydrodynamic
radius with self-complementary sequences self-assemble into fcc
lattice, while CsCl or bcc lattices are obtained by using two
different NPs with complementary DNA sequences
\cite{Macfarlane2013}, where the hydrodynamic radius can be
measured by techniques, such as fast Brownian motion analysis
\cite{Ueberschar2011}. It has been found that DNA NPs with equal
hydrodynamic radii will maximize the number of hybridization
events between neighboring colloids. Additionally, the size ratio
(expressed through the ratio of the hydrodynamic radii) and the
DNA linker ratio (the ratio of the number of DNA linkers) between
two NPs control the thermodynamically favored crystal structure
\cite{Macfarlane2011}. In particular, the size ratio determines
the packing of particles in the crystal and its stability, while
the linker ratio determines the stability of the crystal.
Therefore, two systems with the same size and linker ratio exhibit
the same ``thermodynamic product'' \cite{Macfarlane2011,
Macfarlane2013}. The final crystal structure is the one that
maximizes the DNA sequence-specific hybridization interactions. It
has been also shown that similar structures can be produced by
varying the rate at which individual DNA linkers dehybridize and
subsequently hybridize \cite{Macfarlane2011}. These conclusions
have been reproduced by using a simple model based on the
assumption that DNA sticky ends should physically contact each other
to hybridize and that any sticky ends coming into contact will
eventually form a DNA duplex. Hence, this simple model can be used
as a tool to inform the self-assembly of DNA-functionalized
spherical NPs \cite{Macfarlane2011}. The work of Macfarlane
et al.  \cite{Macfarlane2011,Macfarlane2010, Macfarlane2013}
highlights the key experimental achievements and conclusions
regarding the self-assembly of spherical colloids via DNA-mediated
interactions \cite{Macfarlane2011, Macfarlane2013}.

Conjugation and characterization methods, three representative
self-assembly strategies of the self-assembly of DNA NPs into
super-lattices \cite{Mazid2014}, and design principles for
plasmonic nano-structures \cite{Jones2011b,Tan2011,Fan2011,Baker2010} have
been recently discussed. For example, how DNA has been used to
build finite-number assemblies (plasmonic molecules), molecularly
driven plasmonic switches \cite{Sebba2008}, regularly
spaced NP chains (plasmonic polymers), and extended two- (2D) and
three-dimensional (3D) ordered arrays (plasmonic crystals)
\cite{Tan2011,Soto2002}, as well as DNA patchy particles
\cite{Feng2013} and oligonucleotide capsules \cite{Johnsson2007},
and the formation and stability of surface-tethered
DNA gold-dendron NPs \cite{Hussain2003} have been discussed. Some
work has dealt with the plasmid-templated shape control of condensed
DNA block copolymer NPs \cite{Jiang2012}. Of particular interest has also been the integration of self-assembly with top-down lithography
\cite{Hung2010} for nano-patterning of materials onto substrates
\cite{Singh2011}.

Geerts  et al. \cite{Geerts2010} have reviewed various
aspects of DNA-functionalized NPs self-assembly, such as
the synthesis of DNA NPs nano-complexes, their physical
properties and a range of applications for nano- and micron-sized
colloids either in two- or three-strand systems. At the same time, in
two-strand systems, the DNA strands belonging to different NPs
hybridize directly to each other,  in three-strand systems a third
DNA linker is required for binding the DNA strands belonging to
different NPs. The effect of the number of base pair matches,
length, DNA grafting density on the colloidal surface, ionic
strength of the surrounding medium and effects of pH and solvent
polarity, etc., are parameters discussed in this work.
In particular, understanding and controlling DNA-mediated
interactions between colloidal particles is a key to the
self-assembly process. A number of factors contributing to these
interactions have been investigated by experiments, theoretical
studies and computer simulations designed to investigate
interactions and aggregation/melting behavior of
DNA-functionalized NPs \cite{DiMichele2013}. Moreover, the
clustering of DNA-functionalized colloids that occurs during the
self-assembly shares similarities with percolation phenomena
\cite{Geerts2008}. Geerts  et al. \cite{Geerts2010} have also
discussed 1D, 2D and 3D assemblies of DNA-functionalized colloids,
where for the 3D assemblies an account of micron- and nano-sized
NPs is given. Additionally, a discussion on the broad range of
applications of such systems is provided focusing on applications
such as the detection of small molecules with DNA sequences
(bio-recognition, e.g., within the cell), or DNA/RNA delivery
into cells \cite{Geerts2010}.

The self-assembly of DNA-functionalized colloids has been
interesting in a number of different contexts apart from the
self-assembly. For example, recent advances in the synthesis of
bio-molecule NP/nanorod hybrid systems applications in the
generation of 2D and 3D ordered structures in solutions and on
surfaces have been a large area of research \cite{Katz2004}. The
potential use of bacteria as a novel biotechnology to facilitate
the production of NPs is also an important application, while
emphasis has been also put on bio-molecule-NP assemblies for
bio-analytical applications and for the fabrication of
bio-electronic devices \cite{Katz2004}. Examples of such
applications are sensors, bio-molecule-functionalized magnetic
particles, and bio-molecule-based nano-circuity and viral
templates \cite{Katz2004}. Also, composite assemblies of nucleic
acids, proteins and NPs have been discussed in the literature
\cite{Katz2004}, as well as the protection and promotion of UV
radiation-induced liposome leakage via DNA-directed assembly with
gold (Au) NPs \cite{Dave2011,Dave2011b}.

Heuer  et al. \cite{Heuer2013} highlighted the importance of
Au functionalized conjugates for biomedical applications (e.g.,
biological sensing \cite{Peled2010}) while suggesting a
copper-free click chemistry for the programmed ligation of
DNA-functionalized Au-NP \cite{Heuer2013a}. The focus has been
on the invention and applications of ``nano-flares''
\cite{Chen2013,Zhang2012,Seferos2007}, which carry reporter
duplexes being released after binding. This signal provides the
ability to detect specific endo-cellular targets such as mRNAs,
microRNAs, or ions and small molecules in real time. Their
properties, stability, cellular uptake and cytotoxicity have been
the main focus for such systems. However, the design of many more
highly multiplexed systems with improved targeting properties are
to be expected in the near future. Sokolova  et al.
\cite{Sokolova2008} discussed the transfection of cell with
DNA NPs and methods for gene transfer into living cells, while
Rosi  et al. \cite{Rosi2005,Rosi2006} focused on the use of
these nano-structures in bio-diagnostics, for example electrical
and electrochemical detection, magnetic relaxation detection, and
nano-wire and nano-tube-based detection. Another application
relates to the colorimetric recognition of DNA intercalators with
unmodified Au-NPs \cite{Xin2009}.

Molecular dynamics simulations and theoretical methods related to
the self-assembly of DNA-functionalized NPs have been briefly
discussed previously \cite{Perez2011, Li2014, Lee2010}. A recent
review has summarized the main theoretical and simulation models
in the context of recent experimental progress, highlighting the
advantages and disadvantages of these models and key findings
\cite{Zhang2015}. Moreover, the crystalline structures obtained so
far by experiment have been summarized \cite{Zhang2015}. Despite
the above discussion, a complete presentation of the theoretical
and simulation contribution to the field is still lacking. This
review intends to fill this gap and aims at presenting key
experimental findings and an overview of theoretical and
simulation studies that have contributed fundamentally to
understanding the self-assembly of DNA-functionalized NPs.

\section{Experiments}

\subsection{Synthesis and experimental methodology}

A number of DNA-functionalized NPs-based structures can be formed
by tuning independently the NPs size, shape, and composition.
However, a general strategy for the synthesis of
DNA-functionalized NPs has been discussed only recently
\cite{Zhang2013}, where also NPs with high-density shells of
nucleic acids can be built. Specific coating technology on the NP
surface has been implemented in order to realize this general
synthetic strategy of DNA-based self-assembly. The creation of
heterogeneous NP super-lattices is based on carboxylic-based
conjugation \cite{Zhang2013a}. These heterogeneous materials have
novel optical and field-responsive properties, which are achieved
by a careful selection of NP characteristics, such as size, shape,
and the number of DNA strands per particle and DNA flexibility
\cite{Zhang2013a, Macfarlane2013}. An important advance in the
synthesis of DNA-coated NPs is the synthesis of solid colloids
with surface-mobile DNA linkers \cite{vanderMeulen2013}. In this
case, the DNA linkers are fully mobile on the surface of the NP,
and the temperature range for equilibrium self-assembly is much
broader than in the case of immobile linkers. By increasing the
temperature window for melting, a better control of the
self-assembly process can be achieved, because fast crystallization
leading to kinetically trapped nano-structures can possibly be
avoided. Moreover, linkers accumulate at the NPs below the melting
temperature, thus increasing the number of hybridization events
and the stability of the final assembly structure
\cite{vanderMeulen2013}. A recent computer simulation study has
drawn similar conclusions \cite{Uberti2014}.

Although initially the self-assembly was based on Au NPs,
self-assembly based on silver NPs conjugates by using triple
cyclic disulfide moieties \cite{Lee2007} or metal oxide NPs
\cite{Han2012} has been also reported. DNA-induced size-selective
separation of mixtures of Au-NPs is also possible \cite{Lee2006}
by using the co-operative binding properties of DNA-functionalized
NPs. This is based on the fact that the melting temperature of the
NPs aggregates depends on the size of the NPs \cite{Yang2006}.
Then, a combination of sedimentation and centrifugation techniques
can be used to separate larger NPs from smaller ones
\cite{Lee2006}. This method of separation of NPs according to
their size has also been demonstrated in the case of a ternary
mixture \cite{Lee2006}. Controlled structural evolution of large
silver NPs and their DNA-mediated bimetallic reversible assemblies \cite{Chen2008}
has been recently studied by Kim  et al. \cite{Kim2012}, as
well as the synthesis and reversible assembly of DNA Au-NP cluster
conjugates \cite{Kim2009a}.

It has been recently shown that in binary mixtures of sub-micron
colloids, certain crystals can be obtained from parent crystals by
diffusionless transformations, analogous to the Martensitic
transformation of iron \cite{Casey2012}. This also provides a possibility of accessing certain structures by these parent
crystals, which otherwise would be kinetically unaccessible,
though this structure has a lower energy. An example of this
approach is the formation of a CuAu-I structure from a CsCl
crystal structure \cite{Casey2012}. This work underlines the fact
that structural transformation of atomic solids may also occur in
DNA-linked NPs. An exciting feature of DNA NP assembly structures
is that the NP crystals form reversibly during heating and cooling
cycles. For example, the bcc lattice structure is temperature
tunable and non-compact with particles occupying only a small
fraction of the volume of the unit cell
\cite{Nykypanchuk2008,Dillenback2006}. The compression properties
of DNA NPs assemblies have been studied by measuring their
response to the applied osmotic pressure showing that lattices
based on these materials are super-compressible and capable of transforming between different structures \cite{Srivastava2013}. This
explains the capability of these systems of transforming between
different structures. The lattices of nano-particles
interconnected with DNA, exhibit an isotropic transformation under
compression with a remarkably strong decrease of the lattice
constant, up to a factor of about 1.8, corresponding to more than
$80\%$ of the volume reduction \cite{Srivastava2013}. The force
field extracted from experimental data can be well described by a
theoretical model that takes into account the confinement of DNA
chains in the interstitial regions. It has been shown that
compression properties of these systems can be tuned via DNA
molecular design \cite{Srivastava2013}. Moreover, the structural
plasticity of bio-molecules and the reversibility of their
interactions can also be used to make nano-structures that are
dynamic, reconfigurable, and responsive. The inter-particle
distances in the super-lattices and clusters can be modified,
while preserving structural integrity by adding molecular stimuli
(simple DNA strands) after the self-assembly processes have been
completed. Both systems were found to switch between two distinct
rigid states, but a transition to a flexible state configuration
within a super-lattice has shown a significant hysteresis
\cite{Maye2010}.

The crystal growth depends on the length of DNA linker, the
solution temperature, and the self-complementarity of DNA strands and can be
compared to the case where non-self-complementary DNA linker
strands are used \cite{Macfarlane2009}. A three-step process for
crystal growth has been suggested, i.e., an initial ``random
binding'', resulting in disorder Au-NP aggregates, a localized
reorganization, and a subsequent growth of crystalline domain
size, where the resulting crystals are well-ordered at all subsequent
stages of growth \cite{Macfarlane2009}. A stepwise growth process
to systematically study and control the evolution of a
bcc crystalline thin-film of
DNA-functionalized NPs on a complementary DNA substrate has been
recently presented \cite{Senesi2013}. This crystal growth on a
substrate has been examined as a function of the temperature, the
number of layers, as well as the substrate-particle bonding
interactions. In this case, the interfacial energy between various
crystal planes and the substrate were tuned by the DNA
interactions controlling the crystal orientation and size in a
stepwise fashion using chemically programmable attractive forces.
This is a unique approach, since the prior studies involving
super-lattice assembly typically relied on repulsive interactions
between particles to dictate the structure, and those that relied on
attractive forces (e.g. ionic systems) still maintained repulsive
particle-substrate interactions \cite{Senesi2013}. Indeed, to
date, 17 unique symmetries have been realized and over 100 unique
crystal structures have been synthesized, all of which conform to
a key hypothesis: the obtained assembly structures maximize the total
number of hybridized DNA interconnects between particles
\cite{Senesi2013}.

The number of possible lattices for synthetically programmable NP
super-lattices  can be increased by using a hollow 3D spacer
approach \cite{Auyeung2012}. In this approach, some NPs within a
binary lattice can be replaced with ``spacer'' entities that are
designed in such a way as to replace the behavior of the missing NPs.
For example, the NPs of one type are omitted by the structure
without affecting the positions of the other NPs. In this way,
super-lattices with five distinct symmetries have been created,
including the one that has never been observed before in any
crystalline material \cite{Auyeung2012}, while the nano-meter
precision in lattice parameter tunability has been maintained
\cite{Auyeung2012}. Although hexagonally close-packed arrays of
NPs are quite commonly made by different assembly strategies,
simple hexagonal lattices of NPs have been more difficult to
create, because close-packed structures are often energetically
more favorable. Furthermore, a graphite arrangement of spherical
NPs has never been made by any assembly method. However, by
applying the 3D spacer approach to the AB$_{2}$ lattice, both of
these lattices are easily synthesized and exhibit a crystalline
order with micro-meter-sized domains \cite{Auyeung2012}. Hollow
triangular gold-silver alloy nano-boxes have been also reported
\cite{Keegan2013}, a process which is reversible upon heating with
sharp melting transitions, which has been monitored at wavelengths
throughout the visible and near-IR.

\subsection{Phase behavior and influence of various parameters}

An elegant way to regulate the range and magnitude of forces
between pairs of different particles is based on the control of
the specific DNA linkages between colloidal particles
self-assembly through the combination with polymer brushes
\cite{Valignat2005}. Such self-assembly has been found to be a
reversible process. The key of reversibility is not to allow the
particles to be very close, where van der Waals forces become
important. This is why the presence of adsorbed
polymers is required \cite{Valignat2005}. In fact,
DNA-functionalized Au NPs in macromolecularly crowded polymer
solutions have been recently reported \cite{Shin2012}.

The inter-particle distance of fcc structures can be tuned by
increasing the DNA linker length or the total DNA length
\cite{Hill2008}. Structures containing less than $0.52\%$ of
inorganic material have been synthesized, enabling the tuning of
the unit cell lattice parameters within a large range of values
\cite{Hill2008}. Moreover, the stability of the structures depends on
the sequence of the DNA bases
\cite{Storhoff2002, Doyen2013}.
Namely, different DNA sequences will guide the self-assembly of
these NPs into different structures \cite{Park2008}. In binary
mixtures, the degree of binding between complementary beads
depends also on the number of matching pairs and the ionic
strength of the solution \cite{Milam2003}. A large variety of
colloidal structures, such as chains of alternating large and
small particles for a large range of particle number ratios and
volume fractions have been investigated \cite{Milam2003}. Also, the
plasmon coupling strength depends strongly on the length of the
DNA connecting Au-NPs \cite{Lubitz2011}. The choice of the DNA
linking molecules and the absence or presence of a non-bonding
single-base flexor can be adjusted, so that Au NPs assemble into
micro-meter-sized fcc or bcc crystal structures. These results
demonstrate that a single-component system can be directed to form
different structures \cite{Park2008}. Moreover, particles with a
poly-dispersity of $20\%$ do not form well-defined crystalline
assemblies. Crystalline structures usually require a
poly-dispersity of NPs below $10\%$ \cite{Park2008}.

For 3D hybridization with polyvalent DNA Au NP conjugates
\cite{Hurst2008}, it has been found that there is a minimum number of
base pairings necessary to stabilize DNA Au NP aggregates as a
function of salt concentration for particles between 15 and 150~nm
in diameter. Sequences containing a single base pair interaction
are capable of effecting hybridization between 150~nm DNA Au NPs.
While traditional DNA hybridization involves two strands
interacting in one dimension (1D, Z), hybridization in the context
of an aggregate of polyvalent DNA Au NP conjugates occurs in 3D
making NP assembly possible with three or fewer base pairings per
DNA strand. These studies enable the comparison of the stability
of duplex DNA free in solution and bound to the NP surface. It has
been found that 4--8, 6--19, or 8--33 additional DNA bases must be
added to free duplex DNA to achieve melting temperatures
equivalent to hybridized systems formed from 15, 60, or 150~nm
DNA Au NPs, respectively \cite{Hurst2008}. In addition, the
equilibrium binding constant ($K_{\rm eq}$) for 15~nm DNA Au NPs
(three base pairs) is three orders of magnitude higher than the
$K_{\rm eq}$ of the corresponding NP free system \cite{Hurst2008}.

Xiong  et al. \cite{Xiong2008,Xiong2009} provided the first
experimental phase diagram of the NPs assembled by DNA linkers in
terms of nominal DNA linker/Au-NP ratio $r$ and volume fraction $\eta$,
where the boundary between disordered and bcc structures has been
determined. The formation of a crystalline bcc phase was observed
for a limited range of linker lengths, while the number of linkers
per particle controlled the onset of system crystallization. The
effect of linkage defects on the crystalline structure was also
examined. Note that these results were corroborated later
by computer simulations \cite{Knorowski2011}.

Although a broad range of different non-compact structures has
been created over the last years \cite{Cigler2010}, a
single-component NP assembly of a diamond structure has not been
reported. By a delicate tuning of the particle interactions,
binary diamond-like crystalline structures have been fabricated
\cite{Cigler2010} though. For example, NaTl-type non-compact
structures have been reported, using surface-modified Q$\beta$
phage capsid particles and Au-NPs of similar radii. This structure
has an inorganic component and an organic virus-like protein NP
interpenetrating diamond lattices. This is an example where
particles of different chemical nature can be combined after
a proper surface processing, without phase separation taking place.
However, protein-based structures had been discussed already
before \cite{Park2001}, in addition to the inorganic NPs
\cite{Park2004}. NP-DNA conjugates bearing a specific number of
short DNA strands by enzymatic manipulation of NP-bound DNA have
been also reported \cite{Qin2005}, as well as the self-assembly of
icosahedral virus particles organized by attached
oligo-nucleotides \cite{Strable2004}, and binary heterogeneous
super-lattices assembled from quantum dots and Au-NPs with DNA
\cite{Sun2011}, or assembling colloidal clusters using crystalline
templates and reprogrammable DNA interactions \cite{McGinley2013}.

\subsection{Micron-sized nano-particles}

Micron-sized latex particles with single-stranded DNA grafted to
the surface have been used as a model system to study DNA mediated
interactions \cite{Nykypanchuk2007}. The presence of DNA strands
allows for the tuning of the interactions between NPs by combining
hybridizing linker DNA with non-hybridizing neutral DNA, which
leads to a gradual change of the assembly rate. The effect of
linker/neutral DNA ratios on particle assembly kinetics and
aggregate morphology has been experimentally investigated for a
range of ionic strengths. The conditions for controlling various
assembly morphologies have been identified, and the involved
attractive and repulsive interactions have been described and
explained for the proposed approach, also accounting for further
theoretical considerations \cite{Nykypanchuk2007}.

Kim  et al. \cite{Kim2006,Kim2009} have reported on the first
DNA-mediated micrometer sized colloidal crystals by using
polystyrene particles, discussing preparation strategies of these
colloids in order to obtain such crystals. Thermodynamic
properties of the self-assembly of these crystals are also
predicted through a simple thermodynamic model. Possible ways of
obtaining more ordered structures by studying the growth kinetics
and phase behavior of the self-assembly of DNA-functionalized
colloids have been also considered \cite{Kim2009}. Various aspects
of the selective, controllable, and reversible aggregation of
polystyrene Latex micro-spheres via DNA hybridization have been
discussed \cite{Rogers2005}. These systems were found to have
interesting kinetics and non-exponential binding behavior
\cite{Rogers2013}.

The first direct measurements of DNA-induced interactions between
a pair of micron-sized polymer colloids have been conducted by
Biancaniello  et al. \cite{Biancaniello2005} by using an
optical tweezer method. These results can be modelled using
standard statistical mechanics methods. It has been
shown that micron-sized spherical colloids have binding dynamics
of a power-law in annealing and crystallization
processes. Additionally, the phase behavior of DNA-mediated
micro-sphere suspensions was determined as a function of salt and
oligonucleotide concentration \cite{Biancaniello2007}. In this
case, fluid phases, aggregate phases, or mixed fluid phases with
aggregates have been observed, while increasing salt or
hybridizing events favors the formation of aggregates. Forces of
interaction between DNA-grafted colloids via optical tweezer
measurements have been also reported elsewhere \cite{Kegler2007},
in quantitative agreement with a proposed theoretical model
\cite{Kegler2008}. The specific interaction of DNA-functionalized
polymer colloids by using DNA linkers in solution has been also
reported \cite{Chollakup2010}. Based on such measurements, a
Monte Carlo (MC) scheme to describe colloidal crystallization has
been implemented \cite{Scarlett2010,Scarlett2011}.

Switchable self-protected attractions in DNA-functionalized
spherical colloids are also very interesting \cite{Leunissen2009}.
For example, single-stranded DNA can fold creating secondary
structures, such as loops and hairpins
\cite{Feng2012,Claridge2005,Maye2006,Maye2007}, which influence the
self-assembly. For micron-sized particles, these structures provide a
control over inter-particle binding strength and association
kinetics. By heating and cooling the system, one can mediate these
effects. Such topological interactions may lead to new materials
and phenomena, like particles strung on necklaces, confined
motions on designed contours and surfaces, or even colloidal arrangement in the shape of ``Olympic
gels'' \cite{Feng2012}. Also, polyvalent nucleic acid
nano-structures (PNANs), which are composed of only cross-linked
and oriented nucleic acids, are of great interest. Apart from the
self-assembly properties, these particles are capable of effecting
high cellular uptake and gene regulation without the need of a
cationic polymer co-carrier \cite{Cutler2011}.

\subsection{Anisotropic DNA-functionalized nanoparticles}

Some structures cannot be fabricated by using only spherical
colloids in all dimensionalities. Hence, directional bonding
interactions which can be tuned by the different geometry of the
NP have been used for a better control over the directed
self-assembly of such DNA NPs. One advantage of using particle
anisotropy is the fact that a greater surface contact favors a
higher number of hybridizations. Moreover, the effective local
concentration of terminal DNA nucleotides that mediates
hybridization increases, while conformational stresses imposed on
nano-particle-bound ligands participating in interactions between
curved surfaces might be smaller \cite{Jones2011, Macfarlane2013}.
DNA NPs with core particles of different shape have been
synthesized in a way to study the effect of the core-core particle
interactions, affecting super-lattice dimensionality,
crystallographic symmetry and phase behavior
\cite{Jones2010,Jones2011}.

An example of 1D structures is the self-assembly of nano-rods
\cite{Dujardin2001}, where again the most stable structure is the
one that maximizes hybridization events between DNA sticky ends
\cite{Macfarlane2011}. Moreover, for octahedron NPs, one might
obtain fcc or bcc structures depending on the length of the DNA
strands. However, previous work \cite{Jones2010,Jones2011} has
obtained self-assembly aggregates rather than real long-range
crystal structures indicating the difficulty in obtaining
long-range crystalline structures. Recently, the use of DNA-driven
assembled phospholipid nano-discs as a scaffold for Au-NP
patterning has been discussed \cite{Geerts2013}. A facile and
efficient preparation of anisotropic DNA-functionalized Au NPs and
their assembly has been reported recently \cite{Tan2013}. This
one-step solution-based method has been used to prepare
anisotropic DNA-functionalized Au-NPs with $96\%$ yield and with
high DNA density. Well-defined nano-assemblies using Au NPs having
a specific number of DNA strands have been also discussed
\cite{Qin2008}. Periodic square-like Au NP arrays templated by
self-assembled 2D DNA nano-grids on a surface have been also
reported in the literature \cite{Zhang2006}, as well as
multiple-nano-component arrays by 2D DNA scaffolding
\cite{Pinto2005}. Triangular gold nano-prisms \cite{Millstone2008}
in the presence of attractive depletion forces and repulsive
electrostatic forces assemble into equilibrium 1D lamellar
crystals in solution with inter-particle spacings greater than
four times the thickness of the nano-prisms \cite{Young2012}.
Experimental and theoretical studies reveal that the anomalously
large spacings of the lamellar morphologies are due to the balance
between depletion and electrostatic interactions \cite{Young2012}.

DNA-directed assembly of asymmetric nano-clusters using Janus NPs
has been also investigated \cite{Maye2009,Xing2012}. Curvature effects in
DNA Au-NP conjugates are notably important, due to the asymmetry of
the NP \cite{Xu2006,Cederquist2009}. Particle curvature also plays an
important role in controlling the DNA surface density
\cite{Cederquist2009}. Based on such an approach, there is a
method to predict DNA packing on non-spherical particles. An
example of usage has been provided for Au nano-rods
\cite{Cederquist2009}. Linear micro-structures in DNA nano-rod
self-assembly have been reported recently \cite{Vial2013}. Fibers,
tubules, and ribbons are typical 1D nano-structures that require
anisotropic interactions that also emerge from the geometry of the
NP, which affects the underlying inter-particle interactions
during the structure formation. For systems of DNA-functionalized
nano-rods, the resulting structures are 1D ladder-like meso-scale
ribbons with a side-by-side rod arrangement. Moreover, the DNA can
bind reversibly, facilitating in this way the formation of
hierarchical assemblies with time. Linear structures of
alternating rods and spheres have been also obtained, implying the
generic applicability of the mechanism for nano-scale objects
interacting via flexible multiple linkers \cite{Vial2013}.

\subsection{Self-assembly in one- and two-dimensional geometries}

DNA-based self-assembly of NPs has been discussed in the context
of planar or other geometries, as, for example, in the case of
DNA-mediated 2D colloidal crystallization above different
attractive surfaces \cite{Jahn2010}. Cheng  et al.
\cite{Cheng2009} reported the first DNA-based strategy for
mono-layered free-standing NP super-lattices or very thin films.
In this case, discrete free-standing super-lattice sheets can be
obtained without the requirement of Watson Crick's specific binding in
which inter-particle spacing occurs within the structrure, leading to
the formation of plasmonic and mechanical properties that can be
rationally tuned by adjusting the DNA length  \cite{Lan2013}.
%
%
Moreover, thin standing films based on
DNA-coated colloids with double-stranded DNA have been reported
\cite{Geerts2010a}. Such colloids can self-assemble into unique
crystalline mono-layers, suspended at a distance of several
colloidal diameters above a weakly adsorbing substrate, without
a dependence on DNA hybridization. The advantage of such systems is
that they can be prepared in one location and then be used in a
different application somewhere else. This may be a way for the
assembly of multi-component, layered colloidal crystals
\cite{Geerts2010a}. Moreover, responsive multi-domain
free-standing films of spherical Au-NPs assembled by DNA-directed
layer-by-layer approach have been also studied recently
\cite{Estephan2013}. Large-area spatially ordered arrays of Au-NP
crystals directed by lithographically confined DNA origami have
been also synthesized \cite{Hung2010a} and 5~nm size Au-NPs
arranged into long-range ordered 2D arrays onto lithographically
patterned substrates \cite{Hung2010a}. Additionally, the first
DNA-mediated formation of supra-molecular mono- and multi-layered
NP structures has been presented \cite{Taton2000}, as well as DNA
and DNAzyme-mediated 2D colloidal assembly \cite{Shyr2008}.

\section{Computer simulations}

Various simulation models for DNA chains have been contrived over
the last years
\cite{DeMille2011,Kenward2009,Drukker2001,Knotts2007,Sambriski2009,Lequieu2015,Linak2011,Mladek2013,Ding2014,Sales2005,Starr2006,Theodorakis2013}.
As it usually happens with these models, they aim at
reproducing only the properties of interest of the system. Due to
the large system sizes and time scales required in the study of
DNA or DNA-coated NPs, mainly coarse-grained models are capable of
describing relevant phenomena or bespoke theoretical models, such as the self-assembly of
DNA-functionalized NPs.

\subsection{Coarse-grained models}

\begin{figure}[ht]
\centerline{\includegraphics[width=0.65\textwidth]{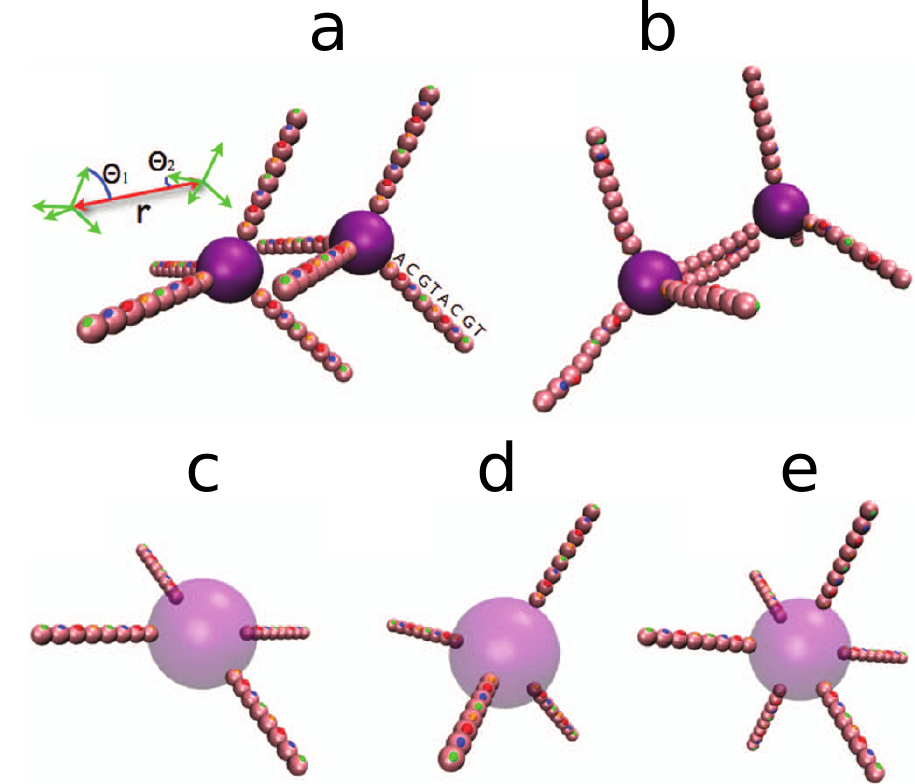}}
\caption{(Color online) Schematic representation of the model of Starr and
Sciortino \cite{Starr2006}. This model considers only a specific
DNA sequence (specified in panel a) which leads to the
hybridization shown in panel (b). This model can assume different
DNA lengths, NPs sizes and grafting geometries (c)--(e). The
effective potential between two NPs modelled in this way depends
on the angles $\theta_{1}$ and $\theta_{2}$, and the distance $r$
between the NPs. Further details on how to calculate the effective
potential can be found in \cite{Theodorakis2013}. Figure
adapted from \cite{Theodorakis2013} after permission.}
\label{fig:1}
\end{figure}

Starr and Sciortino \cite{Starr2006} have proposed a simple model
for the study of the self-assembly of DNA-functionalized NPs
(figure~\ref{fig:1}). This model has been used for short DNA strands
(six to twelve bases) and specific DNA sequences. The effect of
several parameters has been explored, for example, the number of
bases per strand, the number of linking bases and the number of
spacer bases on the stability of crystal states, as well as the
effect of using a single linking NP type versus a binary linking
system. Additionally, the free energy, entropy, and melting point
for bcc and fcc lattices have been explicitly calculated. While
binary systems preferably form bcc lattices, melts of single NP
type with chains of self-complementary basis form the most stable
fcc crystals. A way to maximize a crystal stability based on the
heat of fusion between crystal and amorphous phases growth has
been discussed \cite{Vargas2011}. The stability of bcc and fcc
crystals formed by DNA-linked NPs has been further investigated
\cite{Padovan2011} as a function of the number of attached strands
to the NP surface, the size of the NP core, and the rigidity of
the strand attachment. It has been found for this model that
relaxing the condition of kinetically immobile strands on the
surface of the particles, a slight increase or decrease in the
melting temperature occurs. Larger changes to the melting
temperature result from increasing the number of strands, which
increases the melting temperature, or by increasing the core NP
diameter, which decreases the melting temperature
\cite{Theodorakis2013}. A quantitative description of the melting
temperature has been provided by the model of Starr and Sciortino
\cite{Starr2006,Vargas2011,Theodorakis2013}. This model has been
fully investigated and used in the study of self-assembly process
\cite{Largo2007,Dai2010}. In these studies, a phase diagram for
the NPs as a function of temperature and NPs density has been
provided.

Furthermore, the effective potential between DNA-functionalized
NPs can be obtained enabling the comparison of the phase diagram
of effective particles with the phase diagram of the NPs based on
the full coarse-grained description. Various grafting geometries
have been considered in studying the self-assembly, such as
triangular planar, tetrahedral, square-based pyramid,
triangle-based bipyramid, and octahedral geometries
\cite{Dai2010}. These NPs have short DNA strands with imposed
rigidity and complementary DNA sequences resembling interactions
between patchy colloids.  For all these geometries, but the
octahedral one, there has been found a phase behavior analogous to
gas-liquid transition. For the octahedral geometry, various
crystal structures of similar symmetry occur, depending on the
chemical potential (density) of the system. Then, interpenetrating
networks of DNA-functionalized NPs occur \cite{Dai2010}. In the
following study \cite{Dai2010a}, a two-step crystallization of
DNA-functionalized NPs has been discussed, where the narrow
temperature range that the self-assembly takes place is described
through a two-step process, mediated by a highly connected
amorphous intermediate. Moreover, at lower temperatures, the system
is kinetically trapped hindering crystallization \cite{Dai2010a}.
By using a standard cluster analysis and relevant order parameters
based on spherical harmonics, the structures and their degree of
crystallinity were resolved. The relaxation and crystallization
times of such systems have been also studied \cite{Dai2010a}.
Different amorphous networked phases induced by multiple
liquid-liquid critical points providing a phase diagram of the
temperature versus density have been identified \cite{Hsu2008},
based on effective NPs. This effective potential has been
carefully revised in the following work \cite{Theodorakis2013}.
Different structure factors can describe the interpenetrating
networks as the density of the NPs increases \cite{Hsu2008}.

The equilibrium clustering dynamics and the self-assembly kinetics
can be described on the basis of the Flory-Stockmayer theory
\cite{Flory1953,Stockmayer1943,Hsu2010} finding good agreement
among theory and results obtained from the coarse-grained model.
Considering only two- and three-fold NPs in order to facilitate
the branching between two-fold DNA NPs, percolation and phase
separation at low density have been studied \cite{Hsu2010}.
Furthermore, the kinetics of the self-assembly are investigated in
terms of the bond formation rate as a function of temperature and
time \cite{Hsu2010}.  It has been found that these processes
follow an Arrhenius behavior, where the self-assembly is described
by two ``reaction'' constants. The model of Starr and Sciortino
\cite{Starr2006} has been also used to study the effects of
bidispersity in DNA strand length \cite{Seifpour2013} and effects
of DNA strand sequence and composition
\cite{Seifpour2013,Seifpour2013a}. Moreover, NP dimers linked by
DNA can be used as the building blocks of a multi-scale,
hierarchical assembly, especially when the size of the NPs is
similar \cite{Chi2012}. The structure of NP dimers through a
combination of scattering experiments and molecular simulations
has been recently investigated \cite{Chi2012}. It has been found
that the inter-particle separation within the dimer is mainly
controlled by the number of hybridized DNA strands, their length
and the curvature of the particle, as well as the excluded volume
effects. Without any free parameters, the increase of dimer
separation with increasing temperature can be interpreted as a
result of the change in the number of connecting DNA
\cite{Chi2012}. Recently, it has been shown that tetravalent DNA
nano-stars never crystallize, but they form a fully bonded
equilibrium gel \cite{Rovigatti2014}. The phase diagram of these
organic complexes has been also discussed \cite{Rovigatti2014b}.

\begin{figure}[!ht]
\centerline{\includegraphics[width=0.65\textwidth]{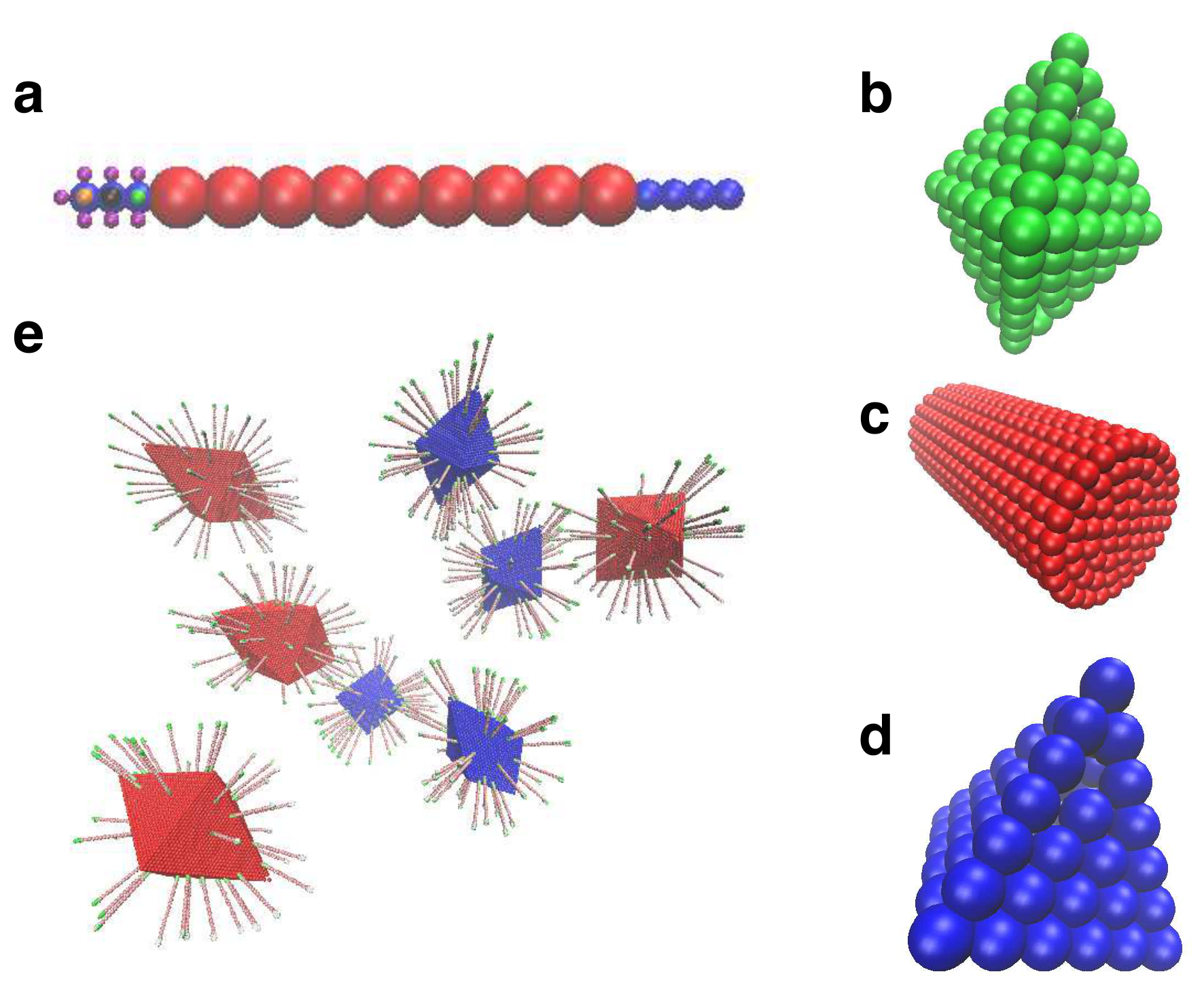}}
\caption{(Color online) Schematic representation of the model by Li  et al.
\cite{Li2013}, as adapted from the original model of Knorowski
 et al. \cite{Knorowski2011}. In panel (a) we present a
typical DNA chain. Blue beads indicate flexible ssDNA, red beads
dsDNA, while green, black, and orange beads represent DNA bases.
The magenta flankening beads have been originally used in \cite{Knorowski2011} to prevent double bonding, in this way
overcoming a drawback of the model of Starr and Sciortino
\cite{Starr2006}. Li  et al. have used spherical particles
for their simulations, but different geometries can be used as
those in panels (b), (c), and (d). In panel (e), we present an
initial configuration for MD or MC simulation of
DNA-functionalized octahedra based on the model of Li  et al.
\cite{Li2013}.} \label{fig:2}
\end{figure}
\looseness=-1A model that is closer to the experimental expectations has been
proposed by Knorowski  et al.
\cite{Knorowski2011,Knorowski2011a,Knorowski2012} and has been
amended by Li  et al. \cite{Li2012} to match the experimental
energy and length scales (figure~\ref{fig:2}). Apart from a number of different
structures and the very good prediction of the experimental phase
diagram as a function of the volume fraction, temperature, and
grafting density, this model is capable of describing the dynamics and
thermodynamics of the self-assembly process. The D-bcc crystalline
phase also exists in their phase diagram. In this case, two different
types of DNA-functionalized NPs occupy the positions of a bcc lattice, but
the type of particles does not have the same pattern across the crystalline structure.
This effect has been previously
overlooked. Moreover, a CsCl-bcc structure is a part of the phase
diagram, while the exact boundary with disordered structures has
been determined, providing the universal dynamics of the
crystallization process \cite{Knorowski2011}. This model has in
many different aspects overcome the problems of the modelling
approach of \cite{Starr2006}, such as the possibility of a
strand to bind to more than one complementary DNA strands.
Recently, this model has been used to study in more detail
the crystallization dynamics, such as gelation, nucleation, and
topological defects \cite{Knorowski2012}. The formation of crystals
involves various dynamic stages of the system, starting from a
liquid mixture to the formation of a gel. Then, the formation of a
crystal containing defects takes place, and finally an equilibrium
structure forms \cite{Knorowski2012}. For these stages, the dynamics of
the process have been discussed in detail \cite{Knorowski2012}. Li
 et al. \cite{Li2012} have actually used this model
\cite{Knorowski2011} to study the phase diagrams as a function of
size ratio and DNA coverage ratio as is done in experiment
\cite{Macfarlane2010}. The authors confirmed fully the
experimental observations \cite{Macfarlane2011, Macfarlane2010}.

In a follow-up study \cite{Li2013}, it was found that the
thermally active hybridization drives the crystallization of
DNA-functionalized NPs. By using Molecular Dynamics (MD)
simulations, the dynamic aspects of the assembly process were
analyzed and the key ingredients to a successful assembly of NP
super-lattices have been identified. The conclusions of the
simulation model are scalable, being able to reproduce the
experimental observations, which have proposed various structures
\cite{Li2013}. Moreover, the optimum percentage of DNA
hybridization events has been determined for different binary
systems, being able to propose suitable linker sequences for
future nano-material design \cite{Li2013}. Recently, this model,
in combination with experiments, has been used to investigate the
DNA-mediated NP crystallization into Wulff polyhedra, where the
use of DNA-functionalized NPs offer a higher control over the growth
process of the crystals \cite{Auyeung2014}.

The pair interactions between DNA-functionalized colloids have
been the main topic in the field \cite{Lequieu2015}, and among other
methods, MC simulations considering NPs with short and stiff DNA
chains have been employed \cite{Leunissen2011}. Such interactions
have been initially measured for planar surfaces corresponding to
large NPs. Based on simulation results, analytical expressions for
the interactions can be obtained, claiming that this is a valid
description for a range of NPs from the nano- to the micro-meter
scales, as well as for a large range of grafting densities.
Moreover, important contributions to the repulsive and attractive
terms of the free energy are coming from purely entropic effects
of the discrete tethered sticky ends. Such entropic contributions
compare well with the hybridization free energy of a free pair of
sticky ends in solutions. Stable gas-liquid separation only occurs
for NPs with radii of a few tens of nanometers, suggesting
different crystallization routes for nano- and micron-sized NPs
\cite{Leunissen2011}. Statistical anisotropy on the distribution
of sticky ends on the NPs' surface leads to a large variation of
the binding strength. This may weaken the reliability of tests
based on the detection of DNA targets in diagnostics. These
results provide a general background for the systems with tethered
binding groups \cite{Guerrini2013}, as for example in the areas of
supra-molecular chemistry or ligand/receptor mediated
bio-recognition \cite{Leunissen2011}. Based on this MC
coarse-grained model, a new strategy to improve the self-assembly
properties of DNA-functionalized colloids has been proposed
\cite{Mognetti2012}. The narrow crystallization temperature window
is due to the sensitivity of DNA-mediated interactions to
temperature and other physical control parameters. By exploiting
the competition between DNA linkages with different nucleotide
sequences, this temperature window can become larger. Thus, the
temperature dependence of the self-assembly process has been
decreased offering a control over a larger temperature range. Such
an approach is also applicable in a realistic situation, where the
control is due to a combinatorial entropy gain, which stems from
the large number of binding partners for each DNA strand.
Experimental work has suggested particular DNA sequences that are
capable of reproducing those effects \cite{Mognetti2012}.

The narrowness of the temperature has been also addressed in
another study by Uberti  et al. \cite{Uberti2012}, since this
narrow temperature window from dilute systems to aggregates
reduces considerably the capability of obtaining defectless crystal
structures, thus restricting the number of possible crystal
structures. A design of DNA-coated colloids which crystallize on
cooling and melt on further cooling has been proposed
\cite{Uberti2012}. Although this is a mere theoretical
model-assumption, it has been argued that realistic situations
that can follow such a phase behavior do exist. While the results
presented refer to binary systems, these concepts also apply in
multi-component systems, opening the way towards the design of
truly complex self-assembling structures based on spherical
colloids \cite{Uberti2012}. The implications of these results have
been commented by Gang \cite{Gang2012}. Latest techniques exploit
the narrow temperature window though, in order to construct
amorphous materials with specific morphology and local separation
between multiple components. In this regard, strategies to direct
the gelation of two-component colloidal mixtures by sequentially
activating selective interactions have been proposed
\cite{DiMichele2013a}. Changes in the structure can be
investigated via MD simulations and experimental methods. This
approach in studying the multi-step kinetics of the self-assembly
process can be extended to the assemblies of multi-component
meso-porous materials with possible applications in hybrid
photovoltaics, photonics, and drug delivery \cite{DiMichele2013a}.

MC simulations combined with histogram re-weighting techniques
have been applied in order to construct the phase diagram of
DNA-functionalized NPs based on a coarse-grained model
\cite{Martinez2010}. This model can predict the phase separation
between a dilute vapor-like phase and a dense phase where NPs are
a part of a liquid-like network based on DNA links. In this case,
there is a non-monotonous dependence of the coexistence pressure on
the temperature, namely the reduced hybridization free energy
\cite{Martinez2010}. This behavior may be attributed to the
crossover between two distinct regimes. Namely, the
hybridization-free-energy driven regime and an entropy-driven
regime. In the former case, there is a ``normal'' vapor-liquid
equilibrium, where the system gains free energy but loses entropy.
In the entropy-driven regime, an increase in entropy occurs due
to the rearrangement of sticky-end bonds in the liquid phase. The
main conclusion of this work is that the system undergoes a phase
separation if the number of DNA-chains per NP is larger than two.
As the number of chains increase, the coexistence region increases
considerably \cite{Martinez2010}. In another MC study
\cite{Martinez2011}, the phase behavior of colloids coated with
long, flexible DNA chains, which can also be achieved
experimentally \cite{Hazarika2007}, has been investigated. In
this case, when the number of DNA strands per NP decreases below a
critical value, the triple point disappears. Also, the condensed
phase coexisting with the vapor remains a liquid. This result may
explain the reason why NPs with small DNA coverage cannot
crystallize in dilute solutions highlighting in this way the
discrete nature of DNA binding \cite{Martinez2011}.

In another coarse-grained approach realized in a MC scheme, a
simplified model was proposed, based on experimental data, which
can describe the phase behavior for DNA-coated colloids
\cite{Mladek2012}. These results have indicated that the
assumption of pair-wise additivity of DNA-mediated interactions
lead to substantial errors in the estimation of  the free energy of the crystal
phase. While these simulations correctly predict the
experimentally obtained crystal structures, despite uncertainties
in the estimates of contour and persistence lengths of ssDNA
chains, they also provide a good estimate of the melting
temperature \cite{Mladek2012}. Hence, further, more accurate
experimental results are required in order to construct a more
accurate coarse-grained approach \cite{Mladek2012}.

The effect of linear shear force on the self-assembly of
DNA-functionalized colloids in two dimensions has been also
investigated \cite{Rabideau2007}. A better sampling of the phase
space is possible in binary systems. From the constructed phase
diagram, a linear dependence of the minimum DNA force necessary
for the line formation on the dimensionless shear rate was found. By
employing a force analysis on the obtained structures, a linear
orientation perpendicular to the axis of the elongations component of
the shear was found, which is the orientation that enables the DNA
attraction to resist to shear \cite{Rabideau2007}. The effect
of internal viscoelastic modes on the Brownian motion of a l-DNA
coated colloid is yet another topic of recent research
\cite{Yanagishima2014}.

Spherical NPs of 1 mm colloids bridged by DNA with 32mm contour
length have been studied in a mixture of two types of particles
with grafted double-stranded l-DNA with short complementary single
stranded overhangs as free binding ends. Confocal microscopy
identified the formation of stable size-limited clusters in which
the two types of particles are in contact. The simulations suggest
that entropic exclusion of the bridging DNA from space between
the nearby particle surfaces hinders the formation of crystals in
milimeter-sized NPs \cite{Schmatko2007}.

\subsection{Further models, theoretical, and numerical approaches}

Specific attraction between NPs can be induced by DNA bridging
creating a large number of periodic structures. Although the
thermodynamics of DNA-functionalized NPs in solution is
well-understood, perfectly suitable coarse-grained models are
still challenging. This difficulty has hindered the design of
complex temperature, sequence and time-dependent interactions
required to describe, for example, materials with highly complex
or multi-component micro-structures or the ability to reconfigure
or self-replicate. A recent study \cite{Dreyfus2010} has provided
high resolution measurements of DNA-mediated interactions in
polystyrene micro-sphere-sized spherical colloids closely related
to the experimental conditions. A numerically tractable model that
quantitatively describes the separation dependence and
temperature-dependent strength of DNA-mediated interactions can be
used, without empirical corrections \cite{Rogers2011}. This model
has been also used to successfully describe the case of grafted
DNA brushes with self-interactions that compete with
the inter-particle formation. This simulation design approach can
propose experimentally realizable nano-structures
\cite{Rogers2011}, despite some criticism of this approach stating
that the agreement with experiment may be fortuitous
\cite{Mognetti2012a}.

Similarly, based on standard thermodynamic arguments, a simple quantitative
model for the reversible association of DNA-coated micron-sized
colloids has been built \cite{Dreyfus2009}. These results were
able to reproduce very well the experimental data regarding the
dependence on temperature and sticky end coverage indicating
clearly the short range of association-dissociation transition of
DNA-coated colloids \cite{Dreyfus2009,Dreyfus2010,Sun2005}. It has
been shown that this transition can be described by a simple model
for different types of DNA and different DNA coverage
\cite{Dreyfus2009}. This model takes into account the feature of
the grafted DNA underlining the significance of the entropy cost
due to DNA confinement onto the surface of the NPs. Here,
an extensive description of the experiments carried
out to validate this simple theoretical model is provided \cite{Dreyfus2009}.
Then, a step-by-step statistical mechanics based model is
developed, which is proven to describe with excellent agreement
the temperature and width of the transition, which are both
essential in the self-assembly of DNA-functionalized colloids
\cite{Dreyfus2010}.

A simple model for the melting and optical properties of DNA
Au-NPs aggregates has been also discussed \cite{Park2003a}. At the
melting temperature, the optical properties of the aggregates
change considerably, a transition that is narrower for larger NPs,
as it has been also discussed elsewhere \cite{Theodorakis2013}.
The extinction coefficient has been computed using the discrete
dipole approximation, and an increasing number of bonds breaks as
the temperature increases. Moreover, the melting temperature
approximately corresponds to the bond percolation threshold
\cite{Park2003}. NPs with radius of 50~nm have been considered,
which compare well with experimentally used NPs. In this MC
scheme, there is a reaction-limited cluster-cluster aggregation
followed by dehybridization of the DNA links. These links form
with a probability that depends on the temperature and the
particle radius. The final structure is also a function of the
number of NPs and the relaxation time. At low temperatures, one
obtains clusters which after a long relaxation time transform into
a compact non-fractal cluster. The optical properties
\cite{Lazarides2000,Lazarides2000a} of these structures can be
estimated by using the discrete dipole approximation. Despite this
structural transformation, the melting transition becomes apparent
at the extinction coefficient at 520~nm wavelength remaining
sharp, while the melting temperature increases with increasing NP
radius as found in the previous model \cite{Park2003a}. However, the
restructuring of NPs increases the link fraction at melting to a
value well above the percolation threshold. The extinction
crossover compares well with experiments on Au/DNA
nano-composites, where the incomplete relaxation is a robust
effect. These results share similarities and may relate to a
possible sol-gel transition in such aggregates \cite{Park2003a}.
Also, photo-switchable DNA-functionalized Au NPs, where
hybridization is affected by the ``photon dose'', can be used for
tuning the DNA-mediated interactions \cite{Yan2012}, as well as
high-energy Al/CuO nano-composites \cite{Severac2012}. The optical
and topological characterization of Au NP dimers linked by a
single DNA double strand has been also discussed
\cite{Busson2011}. Furthermore, connected magnetic micro-particles
with DNA G-quadruplexes have been recently reported in the
literature \cite{Mukundan2013}.

A general theory for predicting the interaction potentials between
DNA-coated colloids has been recently presented
\cite{Varilly2012,Angioletti-Uberti2013}. This theory incorporates the configurational
and combinatorial entropic factors that play a key role in
valence-limited interactions. By enforcing self-consistency,
the modelling can achieve quantitative agreement to detailed MC
calculations. With suitable approximations and in particular
geometries, this theory reduces to the previous successful treatments,
which are now united in a common and extensible framework
\cite{Varilly2012}.

A theory based on an attraction of unlike particles combined with
a type-independent soft-core repulsion has been realized for a
binary system of DNA-functionalized colloids \cite{Tkachenko2002}.
In an experimental system, the repulsion and attraction may be
induced by a DNA solution. Based on a theoretical
treatment\cite{Tkachenko2002}, the system exhibits diverse and
unusual morphologies, such as the diamond lattice and the membrane
phase with an in-line square order. This is the first time, albeit
theoretically, that the diamond structure has been predicted
\cite{Tkachenko2002}. Moreover, the inverse approach has been
discussed, when a particular scheme is proposed for building an
arbitrary desired nano-structure. The conditions for robust
self-assembly of the target structures have been identified, which
includes the minimum number of ``colors'' needed to encode the
inter-particle bonds through the hybridization of DNA pairs.
Furthermore, it was shown that a floppy network with thermal
fluctuations can possess an entropic rigidity and can be described as
a traditional elastic solid. The onset of the entropic rigidity
determines the minimal number of bond types per NP needed to
encode the desired structure. This theoretical model takes
into account thermodynamic considerations providing additional
conditions for the occurrence of the equilibrium structures
\cite{Tkachenko2011}.

Halverson and Tkachenko \cite{Halverson2013} have studied the
formation of complex mesoscopic structures with programmable local
morphology and complex overall shape NPs linked with DNA. The
hybridization of bonds is realized through a harmonic potential
and a Poissonian statistical distribution. These results compare
well with experimental results referring to the melting curve of dimers
of DNA-functionalized NPs, and
ordered structures in the range 5--50~nm were obtained. The key
elements of this design strategy are the symmetry based on the
neighbors being of different type and different sequences of
bonds, the controlled nucleation, and the cooperative binding.
Based on this strategy, a number of complete nano-structures can
be assembled, such as the empire state building
\cite{Halverson2013}. The size limits of self-assembled colloid
structures made using specific short-range interactions have been
also recently discussed \cite{Zeravcic2014}. It has been found
that structures with highly variable shapes made out of dozens of
particles can form with high yield, as long as each particle in
the structure binds only to the particles in their local
environment \cite{Zeravcic2014}. Such an approach allows complete
enumeration of the energy landscape and gives bounds on how large
a colloidal structure can assemble with high yield. For large
structures, the yield can be significant, even with hundreds of
particles \cite{Zeravcic2014}.

A theoretical discussion of a self-assembly scheme which enables
the use of DNA to uniquely encode the composition and structure of
micro- and nano-particle clusters has been presented in a series
of papers by Licata and Tkachenko
\cite{Licata2006,Licata2006a,Licata2008,Licata2009}. Several
important aspects of possible experimental implementation of the
proposed scheme have been discussed. For example, the competition
between different types of clusters in a solution, possible
jamming in an unwanted configuration, and the degeneracy due to
symmetry with respect to particle permutations \cite{Licata2006}.
This methodology has been enabled from the high selectivity of
DNA-mediated interactions between NPs. Based on a simple model and
a more realistic description of the DNA-colloidal system, it is
possible to suppress disordered glassy phases of the systems in
order to obtain self-assembly into perfect crystal structures.
This requires a combination of stretchable inter-particle linkers,
which in a realistic situation may simply correspond to
sufficiently long DNA strands, and, also, a soft repulsive
potential. This approach has been implemented on a large number of
cluster types and is experimentally realizable \cite{Licata2006a}.
The self-assembly for the case of DNA-caged NPs has been also
considered in this work \cite{Licata2009}. The dynamics of models
that contain such key-lock binding interactions, like the
Watson-Crick interactions in the DNA including systems where
antibodies or cell membrane proteins are grafted on NPs can be
described theoretically \cite{Licata2008}. Depending on the
grafting density of the binding groups, two regimes have been
predicted. In the localized regime, at low coverage, the system
exhibits an exponential distribution of particle-departure times.
At higher coverage, there is an interplay between departure
dynamics and particle diffusion. This interplay leads to the phenomena
similar to aging in glassy systems, corresponding to a sharp
increase of the departure times. The diffusive effect is similar
to dispersive transport in disordered semiconductors. Depending on
the interaction parameters, the diffusion behavior ranges from
standard to anomalous, sub-diffusive behavior \cite{Licata2008}.

Hydrodynamics may play an important role in the transformation of
colloid super-lattice crystallites, especially when NPs are
micron-sized \cite{Jenkins2014}. For example, the spontaneous
transformation through annealing from a bcc to a fcc has
been experimentally observed \cite{Casey2012,Nykypanchuk2008}. By
using computer simulations and vibrational mode analysis, one can
discern between different structures that have the same energy,
but hydrodynamic correlations differ. It is the effect of
hydrodynamics that induces anisotropic diffusion resulting in the
formation of different ``child'' structures during the annealing
processes. Therefore, structures based on DNA-functionalized NPs
are not purely determined by thermodynamics, but also require the
consideration of the process and properties which may be the result of
hydrodynamic effects \cite{Jenkins2014}. These phenomena play an
important role given the narrow temperature window of the melting
curve.

DNA-functionalized NPs with mobile linkers have been recently
studied \cite{Uberti2014}, inspired by experimental evidence
\cite{vanderMeulen2013}. This approach combines in a sense the
valency of the otherwise isotropic colloids, due to the mobility
of DNA linkers. By combining theoretical and simulation methods,
this valency of colloids can be tuned by the nonspecific repulsion
between particles. The simulations show that the resulting
effective interactions lead to low-valency colloids in peculiar
``open'' structures, considerably different from those observed in
the case of NPs with immobile DNA linkers \cite{Uberti2014}.

\subsection{Lattice models}

The mapping of microscopic DNA relevant sequences onto the
macroscopic phase behavior has been in the focus of theoretical
and simulation work \cite{Lukatsky2004,Lukatsky2005,Lukatsky2006}.
Such an approach should enhance the efficiency of methods
detecting mutations in specific DNA sequences \cite{Lukatsky2004}.
By using a mean-field model, the surface and bulk dissolution
properties of DNA-functionalized NPs assemblies can be analyzed
\cite{Lukatsky2005,Jin2003a}. A model for multi-component systems
able to form low-symmetry ordered phases has been also proposed
despite the spherical symmetry of the interactions
\cite{Lukatsky2006}. By combining MC simulations with mean-field
theory, the thermodynamic, structural and kinetic aspects of
stripe phases for simple and multi-component lattice model can be
studied \cite{Lukatsky2006}. This lattice model represents a
mixture of spherically symmetric DNA-functionalized spherical NPs
with several species of DNA linkers. However,  non-spherically
ordered structures based on spherical NPs can also be constructed,
where DNA linkers of different lengths can be used to introduce
bonds of different spacing between NPs. The particular example of
a stripe structure has been considered \cite{Lukatsky2006}. In all
cases though, the system was forced to have exactly the same
ground state and the highest degree of connectivity of particles.
Hence, as in experiment \cite{Macfarlane2011}, the most stable
structure is the one that maximizes the number of hybridization
events. However, when the strength of competitive binding is weak,
a four-component system can exhibit two consecutive ordering
transitions, which are due to the re-ordering of superstructures
within the system. Based on those observations, an optimal design
strategy has been proposed for super-structures in multi-component
systems \cite{Lukatsky2006}.

Phase transitions in DNA-linked assemblies can be described by a
decorated-lattice model \cite{Talanquer2006}. In this study, two
types of DNA-linked colloidal mixtures have been considered,
namely, systems with identical NPs functionalized with two
different DNA strands and mixtures of two types of particles with
each one being functionalized with different DNA strands.
Establishing the similarity with Ising models with temperature and
activity dependent effective interactions, various properties for
these mixtures can be derived, such as the evolution of the
dissolution profiles as a function of temperature and number of
grafted DNA strands, in agreement with experimental systems. The
increase of the dissolution temperature with the grafting density
is valid only below a certain threshold. For high grafting
densities, the dissolution temperature becomes a decreasing
function of the DNA coverage. In systems with two different types
of particles, the phase separation involves meta-stable disordered
aggregates when compared to a phase transition of solvent-rich and
ordered phase. The ordering of the colloidal network enhances the
stability of the DNA-linked assembly resulting in the dissolution
of the aggregates at higher temperatures. These results explain
the contrasting evolution of the dissolution temperatures with
increasing probe size in both types of mixtures
\cite{Talanquer2006}.

A 2D lattice model for particles of many different types with
short-range isotropic interactions which are pairwise specific has
been also introduced \cite{Tindemans2010}. The main question
addressed is whether the ground state corresponding to a crystal
structure is unique. Interestingly, it has been shown that it is
sufficient to uniquely determine the ground state for a large
class of crystal structures via MC simulations
\cite{Tindemans2010}.

\subsection{Genetic algorithms}

An interesting approach based on genetic algorithms has been
recently implemented to design DNA-grafted NPs that self-assembly
into desired crystalline structures \cite{Srinivasan2013}. Often,
the self-assembly first completes, and then the obtained
self-assembly crystal structures are determined. By using genetic
algorithms based on an experimentally validated evolutionary
optimization methodology, it is possible to predict, not only
original phase diagrams detailing the regions of known crystals, but
also several new structures (for successful applications of
evolutionary algorithms in soft matter systems with directional
bonding see
\cite{Gottwald2005,Fornleitner2008,Doppelbauer2010,Doppelbauer2012}).
The agreement of these results with the existing experimentally
predicted structures validates the use of this tool for the design
of materials based on this genetic technique
\cite{Srinivasan2013}. Although the traditional approach is very
useful in order to validate the results, genetic algorithms offer
a way of designing crystal structures with desired properties,
determining in this way the desired structure that underlies those
properties. In an example where two types of colloids are grafted
with complementary ssDNA sequences, ordered colloidal crystals
have been obtained. This methodology is generalizable, fast, and
selective, able at the same time to reproduce parameters relevant
to four currently realized crystals and unobserved structures
\cite{Srinivasan2013}.

\subsection{Atomistic simulations}

Atomistic simulations of 2~nm Au truncated octahedral NPs
functionalized with four single-stranded DNA chains have been also
realized \cite{Lee2009,Lee2010}. However, such an approach is
limited to a small number of short DNA chains similarly to the
model of Starr and Sciortino \cite{Starr2006}, which deviates
considerably from an experimentally realized nano-structure.
However, in this study the authors have explored the possibility
of hydrogen bonds between the adsorbed ssDNAs, finding that there
is not a significant amount of hydrogen bonds. This conclusion is
consistent with the observed increase of the melting
temperature \cite{Lee2009,Lee2010}.

Despite the extensive use of coarse-grained models, a million-atom
MD simulation has been used to study the formation of bcc and fcc
super-crystals of DNA-functionalized Au NPs \cite{Ngo2012}. The
bcc crystal contained 2.77 million atoms, whereas the fcc
contained 5.05, with the diameter of the NP being 3~nm. After 10 ns
DNA structures vary smoothly, while the ion distributions around
DNAs can also be estimated. Analysis of these results shows that
ions bind stronger in bcc structures than in fcc structures. This
effect explains the higher melting temperature of bcc structures
and underlines the significance of entropic effects in fcc
structures. In this work, the Young's and bulk moduli of the
obtained super-crystals have been also obtained. Note that the
Poisson ratios for both super-crystals were close to the ideal
value, that is $1/3$ \cite{Ngo2012}.

\subsection{Simulations of anisotropic nanoparticles}

Anisotropic NPs grafted with DNA have recently appeared in the
literature for the case of nano-cubes \cite{Knorowski2014}, where
the phase diagram similarly to the previous work with coated spherical
NPs has been studied \cite{Knorowski2011,Knorowski2011a}. These
nano-cubes are capable of self-assembling into crystalline, liquid
crystalline, rotator, or non-crystalline phases with both
long-range positional and orientational order
\cite{Knorowski2014}.

\section{Combined simulation and experimental studies}

The growth dynamics for DNA-mediated NP crystallization has been
one of the fundamental questions, because during the formation of
crystalline-structure NPs become kinetically trapped. Moreover, it
has been found that the growth of isolated crystallites is
described by the power law $\sim t^{1/2}$ \cite{Dhakal2013}. The
coalescence of these crystallites significantly slows down the growth process
due to the orientational mismatch of the
crystallites. MD simulations show that the misorientation angle
decreases continuously during the coalescence, which is a
signature of a rotation induced coalescence mechanism. The
coarsening of the conformation at the boundary between coalescing
crystallites takes place after the orientation of the crystals.
When nucleotide chains are larger, the dynamics become faster due
to the enhanced surface diffusion, which more effectively reduces
the curvature at the boundary of two super-lattices. These results
have provided fundamental understanding on the relationship
between NP surface chemistry and its crystallite growth and
coalescence \cite{Dhakal2013}.

A quantitative prediction of the associated thermodynamics and
kinetics of DNA-coated particles considering different
functionalization schemes has been discussed by Leunissen et
al. \cite{Leunissen2010}. Preparation, structural, and optical
features of 2D cross-linked DNA/Au NP conjugates have been also
considered in the literature \cite{Noyong2006,Cobbe2003}, as well
as the tuning of ratios, densities and supra-molecular spacing in
bi-functional DNA-modified Au NPs \cite{Diaz2012}. Experimental
results \cite{Leunissen2010} show that the suspension behavior is
very sensitive to the grafting details, such as the length and
flexibility of the DNA strands, as well as the surface coverage.
Therefore, in order to control the assembly process, emphasis
should be put on the structural and dynamical features of the DNA
coatings. By varying the temperature cycle, beads concentration,
and surface coverage, the association-dissociation dynamics and kinetics
can be studied systematically. Moreover, the tethered backbone
segments onto the NPs' surface has been found to play an important
role in the association-dissociation properties, since they affect
the total number of inter-particle bonds and, as a result, the
configurational entropy cost associated with these bonds.
Independently of the tether backbone, self-complementary
palindromic sticky ends can easily form intra-particle hairpins
and loops considerably altering the association behavior. Hairpin
DNA-functionalized Au colloids have found also application in the
imaging of mRNA in living cells \cite{Jayagopal2010}. Such
structures are more important during faster temperature quenches,
lower concentration and lower surface coverage. Other applications
include the design of ``self-protected'' DNA-mediated
interactions, which enable more versatile self-assembly schemes
\cite{Leunissen2010}.

Self-replicating materials of DNA-functionalized colloids are also
interesting \cite{Leunissen2009a}. Such systems exploit the strong
specificity and thermal reversibility for the interactions between
DNA-functionalized NPs.  The fundamental behavior of the
self-replicating process related to the equilibrium and kinetic
aspects of aggregation-dissociation behavior of particles has been
determined via experimental methods. It has also been found that
the dissociation temperature is very sharp, occurring unexpectedly
at low temperatures. When the fraction of the sticky ends becomes
smaller, the dissociation temperature shifts to even lower scales.
The sharpness of the transition and the fraction of sticky ends
are important parameters in this self-replication scheme.
Moreover, a double-stranded backbone prevents unwanted
hybridization events, such as secondary structure formation, which
leads to peculiar aggregation kinetics. This is due to the competition
between inter- and intra-hybridization events. Additionally, by
functionalizing a single particle species with two distinct DNA
sequences, permanent bonds can form \cite{Leunissen2009a}.

The bending dynamics of DNA-linked colloidal particle chains have
been also studied \cite{Li2010} by using a worm-like chain (WLC)
model to represent the grafted DNA strands onto the NPs surface,
analogously to the case of semi-flexible linear polymers. The
persistence length of these grafted DNA chains can be determined
by monitoring the thermal fluctuations of the chains. In this way,
these lengths can be tuned within a large range from 1 to 50~nm,
which corresponds to a linking length of DNA of 75 to 15 bases.
Interestingly, the bending relaxation dynamics of these chains
matches the theoretical predictions, suggesting the use of such a
WLC DNA model approach for both equilibrium and dynamical studies
\cite{Li2010}.

Park  et al. \cite{Park2006} investigated the structure of
NPs aggregates at room temperature by combining theoretical and
experimental methods. Experimental methods involve extinction
spectroscopy measurements and dynamic light scattering, where Au
NPs have sizes between 60 and 80~nm and DNA strands of about 30
base-pairs. Theoretical studies use calculated spectra for models
of the aggregate structures to determine the structure that
matches the observations. Since particles in this study are larger
than particles used previously, there is a higher sensitivity of
spectra to aggregate structures. Combining theoretical and
experimental results in the range where the highest agreement was
found, it was argued that DNA hybridization takes place under
irreversible conditions at room temperature. The model also takes
into account other effects, such as the poly-dispersity in the
size of NPs and lattice defects. The room temperature aggregate is
always a fractal, and a morphological change from fractal towards
compact structures occurs below the melting temperature
\cite{Park2006}.

The kinetics of DNA-coated sticky particles through the
exploration of a number of properties have been studied by Wu
et al. \cite{Wu2013}. The aggregation rate, which depends on the
rate of NPs encounter and the probability that such an encounter
results in the sticking of the particle, has been studied.
Contrary to the previous toy models \cite{Macfarlane2011}, this one
takes into account a possibility of hybridization. By combining
experimental and theoretical techniques, the aggregation times of
micron-sized particles as a function of DNA coverage and salt
concentration have been obtained. A simple model using
reaction-limited kinetics and experimental oligomer hybridization
rates is in very good agreement with the experimental data. The
Coulomb barrier at the nanometer scale retarding DNA hybridization
is a significant controlling factor of the association process.
This model also allows an easy measurement of microscopic
hybridization rates from macroscopic aggregation, thus enabling
the design of complex self-assembly schemes with a controlled
kinetics \cite{Wu2013}.

\section{Synopsis}

To summarize, there has been a great advancement in understanding
the self-assembly of DNA-functionalized NPs. A large number of
crystalline structures can be fabricated and the effects of
various parameters influencing the self-assembly are now
well-understood. During the last years, a wealth of different uses
of DNA-functionalized NPs has been realized on the basis of
different applications. However, the control of self-assembly and
crystallization kinetics remains still an unresolved issue. This
problem leads to kinetically trapped structures, one of the main
reasons behind the difficulty in fabricating long-range stable
crystalline structures and the efficient growth of crystalline
structures on substrates \cite{Macfarlane2013}. Problems in
synthesis of these NPs in a controllable and predictable way is
also a concern \cite{Macfarlane2013}. Additionally, transitioning
DNA-engineered NP super-lattices from solution to the solid state
is challenging \cite{Auyeung2012b}. Steps toward creating dynamic
self-assembly materials, that is materials which are capable of
changing their crystalline structure in a controlled way due to
specific stimuli, have been also taken
\cite{Macfarlane2013,Radha2014}. Recently, dynamically
interchangeable NP super-lattices have been produced via the use
of nucleic acid allosteric effectors \cite{Kim2013}. This approach
has been suggested for fcc, bcc, CsCl and AlB$_2$ crystal
symmetries \cite{Kim2013}. Another open question is the design of
DNA-functionalized NPs that could lead to the formation of
single-component non-compact diamond structures. Although some
theoretical suggestions exist along these lines, this structure
has not been realized experimentally, or by any simulation work.
Moreover, given the large flexibility of DNA, the formation of
single-component diamond structures with small NPs may not be
possible \cite{Rovigatti2014}. An opportunity window could be
opened via the design with NPs of non-spherical shape which has
been the topic of recent experimental \cite{Jones2010,Jones2011}
and simulation work \cite{Knorowski2014}.

\section*{Acknowledgements}
PET, GK and CD acknowledge financial support by the Austrian Research
Foundation (FWF) under Proj. No.~F41 (SFB ViCoM). In addition, GK
acknowledges financial support by the FWF under Proj. No.~P23910--N16.


\ukrainianpart

\title{Самоскупчування DNA-функціоналізованих колоїдів}
\author{П.Е. Теодоракіс\refaddr{label1}, Н.Г. Фітас\refaddr{label2}, Г. Каль\refaddr{label3,label4}, Ч. Деллаго\refaddr{label4,label5}}

\addresses{
\addr{label1} Відділ хімічної інженерії, Емпіріал Коледж Лондон,  SW7 2AZ Лондон, Великобританія
\addr{label2} Центр Прикладних математичних досліджень, Університет Ковентрі, CV1 5FB, Ковентрі, Великобританія
\addr{label3} Інститут теоретичної фізики, Віденський технічний університет, A--1040 Відень, Австрія
\addr{label4} Центр обчислювального матеріалознавстваe (CMS),  A--1040 Відень, Австрія
\addr{label5} Факультет фізики, Університет Відня,  A--1090 Відень, Австрія
}

\makeukrtitle
\begin{abstract}
Колоїдні частинки з прикріпленими однонитковими DNA (ssDNA) ланцюжками, здатні самоскупчуватися у низку різних кристалічних структур, де гібридизація  ланцюжків ssDNA створює зв'язки між колоїдами, стабілізуючи їх структуру. Залежно від геометрії і розміру частинок, густини
прищеплених ланцюжків ssDNA, а також густини та вибору послідовностей DNA, можна виготовляти низку різних кристалічних структур. Однак залишається незрозумілим і потребує інтенсивних досліджень питання, яким чином ці фактори роблять синергетичний внесок у процес самоскупчення DNA-функціоналізованих частинок нано- або мікророзмірів. Більше того, додаткові труднощі для розв'язання даної проблеми полягають у виготовленні далекосяжних структур завдяки кінетичним критичним елементам при самоскупченні. У статті обговорюються найновіші досягнення теорії та експерименту з особливим наголосом на останні дослідження методами комп'ютерного моделювання.
\keywords DNA-функціоналізовані наночастинки, самоскупчення, експеримент, теорія, комп'ютерне моделювання
\end{abstract}

\begin{thebibliography}{100}


\bibitem{Crocker2008}
Crocker J.C., Nature, 2008, \textbf{451}, 528; \doi{10.1038/451528a}.

\bibitem{Travesset2011}
Travesset A., Science, 2011, \textbf{334}, No. 6053, 183; \doi{10.1126/science.1213070}.

\bibitem{Rogers2015} Rogers~W.B.,  Manoharan~V.N., Science, 2015, \textbf{347}, 639; \doi{10.1126/science.1259762}.

\bibitem{Hurst2009}
Hurst S.J., Hill H.D., Macfarlane R.J., Wu J., Dravid V.P., Mirkin C.A., Small,
  2009, \textbf{5}, No.~19, 2156; \doi{10.1002/smll.200900568}.

\bibitem{Hurst2011}
Biomedical Nanotechnology: Methods and Protocols, Hurst S.J. (Ed.),
Springer Protocols, Methods in Molecular Biology Series, Humana Press, New York, 2011.

\bibitem{Zuccheri2011}
DNA Nanotechnology: Methods and Protocols, Zuccheri G., Samori B. (Eds.),
Springer Protocols, Methods in Molecular Biology Series,  Humana Press, New York, 2011.

\bibitem{Heuer2013}
Heuer-Jungemann A., Harimech P.K., Brown T., Kanaras A.G., Nanoscale, 2013,
  \textbf{5}, No.~20, 9503; \doi{10.1039/c3nr03707j}.

\bibitem{Rosi2005}
Rosi N.L., Mirkin C.A., Chem. Rev., 2005, \textbf{105}, No.~4, 1547; \doi{10.1021/cr030067f}.

\bibitem{Sokolova2008}
Sokolova V., Epple M., Angew. Chem. Int. Edit., 2008, \textbf{47}, No.~8, 1382; \doi{10.1002/anie.200703039}.

\bibitem{Tan2011}
Tan S.J., Campolongo M.J., Luo D., Cheng W., Nat. Nanotechnol., 2011,
  \textbf{6}, 268; \doi{10.1038/nnano.2011.49}.

\bibitem{Rosi2006}
Rosi N.L., Giljohann D.A., Thaxton C.S., Lytton-Jean A.K., Han M.S., Mirkin
  C.A., Science, 2006, \textbf{312}, No. 5776, 1027; \doi{10.1126/science.1125559}.

\bibitem{Niemeyer2001}
Niemeyer C.M., Angew. Chem. Int. Edit., 2001, \textbf{40}, 4128; \\
\doi{10.1002/1521-3773(20011119)40:22<4128::AID-ANIE4128>3.0.CO;2-S}.


\bibitem{Macfarlane2014}
Macfarlane R.J., Thaner R.V., Brown K.A., Zhang J., Lee B., Nguyen S.T., Mirkin
  C.A., Proc. Natl. Acad. Sci. USA, 2014, \textbf{111}, No.~42, 14995; \doi{10.1073/pnas.1416489111}.

\bibitem{Alivisatos1996}
Alivisatos A.P., Johnsson K.P., Peng X., Wilson T.E., Loweth C.J., Bruchez~(Jr.)~M.P., Schultz P.G., Nature, 1996, \textbf{382}, 609; \doi{10.1038/382609a0}.

\bibitem{Mirkin1996}
Mirkin C.A., Letsinger R.L., Mucic R.C., Storhoff J.J., Nature, 1996,
  \textbf{382}, 607; \doi{10.1038/382607a0}.

\bibitem{Hurst2008}
Hurst S.J., Hill H.D., Mirkin C.A., J. Am. Chem. Soc., 2008, \textbf{130}, No.~36,
  12192; \doi{10.1021/ja804266j}.

\bibitem{Macfarlane2013}
Macfarlane R.J., O'Brien M.N., Petrosko S.H., A. M.C., Angew. Chem. Int. Edit.,
  2013, \textbf{52}, 5688; \doi{10.1002/anie.201209336}.

\bibitem{Norris2007}
Norris D.J., Nat. Mater., 2007, \textbf{6}, 177; \doi{10.1038/nmat1844}.

\bibitem{Barrow2013}
Barrow S.J., Funston A.M., Wei X., Mulvaney P., Nano Today, 2013, \textbf{8},
  No.~2, 138; \doi{10.1016/j.nantod.2013.02.005}.

\bibitem{Geerts2010}
Geerts N., Eiser E., Soft Matter, 2010, \textbf{6}, No.~19, 4647; \doi{10.1039/c001603a}.

\bibitem{Hung2010}
Hung A.M., Noh H., Cha J.N., Nanoscale, 2010, \textbf{2}, No.~12, 2530; \doi{10.1039/c0nr00430h}.

\bibitem{Katz2004}
Katz E., Willner I., Angew. Chem. Int. Edit., 2004, \textbf{43}, No.~45, 6042; \doi{10.1002/anie.200400651}.

\bibitem{Macfarlane2011}
Macfarlane R.J., Lee B., Jones M.R., Harris N., Schatz G.C., Mirkin C.A.,
  Science, 2011, \textbf{334}, No. 6053, 204; \doi{10.1126/science.1210493}.

\bibitem{Mazid2014}
Mazid R.R., Si K.J., Cheng W., Methods, 2014, \textbf{67}, No.~2, 215; \doi{10.1016/j.ymeth.2014.01.017}.

\bibitem{Storhoff1999}
Storhoff J.J., Mirkin C.A., Chem. Rev., 1999, \textbf{99}, No.~7, 1849; \doi{10.1021/cr970071p}.

\bibitem{Perez2011}
P\'erez A., Luque F.J., Orozco M., Acc. Chem. Res., 2011,
  \textbf{45}, No.~2, 196; \doi{10.1021/ar2001217}.

\bibitem{Li2014}
Li N.K., Kim H.S., Nash J.A., Lim M., Yingling Y.G., Mol. Simulat.,
  2014, \textbf{40}, 777; \doi{10.1080/08927022.2014.913792}.

\bibitem{DiMichele2013}
Di~Michele L., Eiser E., Phys. Chem. Chem. Phys., 2013, \textbf{15}, No.~9, 3115; \doi{10.1039/c3cp43841d}.

\bibitem{Zhang2015}
Zhang X., Wang R., Xue G., Soft Matter, 2015, \doi{10.1039/C4SM02649G}.

\bibitem{Cao2001}
Cao Y., Jin R., Mirkin C.A., J. Am. Chem. Soc., 2001, \textbf{123}, No.~32, 7961; \doi{/10.1021/ja011342n}.

\bibitem{Macfarlane2010}
Macfarlane R.J., Jones M.R., Senesi A.J., Young K.L., Lee B., Wu J., Mirkin
  C.A., Angew. Chem. Int. Edit., 2010, \textbf{49}, No.~27, 4589; \doi{10.1002/anie.201000633}.

\bibitem{Ueberschar2011}
Uebersch\"ar O., Wagner C., Stangner T., Gutsche C., Kremer F., Polymer, 2011,
  \textbf{52}, No.~8, 1829; \doi{10.1016/j.polymer.2011.02.001}.

\bibitem{Jones2011b}
Jones M.R., Osberg K.D., Macfarlane R.J., Langille M.R., Mirkin C.A.,
Chem. Rev., 2011, \textbf{111}, 3736; \doi{10.1021/cr1004452}.

\bibitem{Fan2011}
Fan J.A., He Y., Bao K., Wu C., Bao J., Schade N.B., Manoharan V.N., Shvets G.,
  Nordlander P., Liu D.R., Capasso~F., Nano Lett., 2011, \textbf{11}, No.~11,
  4859; \doi{10.1021/nl203194m}.

\bibitem{Baker2010}
Baker B.A., Milam V.T., Langmuir, 2010, \textbf{26}, No.~12, 9818; \doi{10.1021/la100077f}.

\bibitem{Sebba2008}
Sebba D.S., Mock J.J., Smith D.R., LaBean T.H., Lazarides A.A., Nano Lett.,
  2008, \textbf{8}, No.~7, 1803; \doi{10.1021/nl080029h}.

\bibitem{Soto2002}
Soto C.M., Srinivasan A., Ratna B.R., J. Am. Chem. Soc., 2002, \textbf{124},
  No.~29, 8508; \doi{10.1021/ja017653f}.

\bibitem{Feng2013}
Feng L., Dreyfus R., Sha R., Seeman N.C., Chaikin P.M., Adv. Mater., 2013,
  \textbf{25}, No.~20, 2779; \doi{10.1002/adma.201204864}.

\bibitem{Johnsson2007}
Johnston A.P., Caruso F., Angew. Chem. Int. Edit., 2007, \textbf{46}, No.~15,
  2677; \doi{10.1002/anie.200605135}.

\bibitem{Hussain2003}
Hussain N., Int. J. Pharm., 2003, \textbf{254}, No.~1,
  27; \doi{10.1016/S0378-5173(02)00672-5}.

\bibitem{Jiang2012}
Jiang X., Qu W., Pan D., Ren Y., Williford J.M., Cui H., Luijten E., Mao H.Q.,
  Adv. Mater., 2013, \textbf{25}, No.~2, 227; \doi{10.1002/adma.201202932}.

\bibitem{Singh2011}
Singh A., Eksiri H., Yingling Y.G., J. Polym. Sci. Pt. B-Polym. Phys., 2011, \textbf{49}, No.~22, 1563; \doi{10.1002/polb.22349}.

\bibitem{Geerts2008}
Geerts N., Schmatko T., Eiser E., Langmuir, 2008, \textbf{24}, No.~9, 5118; \doi{10.1021/la7036789}.

\bibitem{Dave2011}
Dave N., Liu J., Adv. Mater., 2011, \textbf{23}, No.~28, 3182; \doi{10.1002/adma.201101086}.

\bibitem{Dave2011b}
Dave N., Liu J., ACS Nano, 2011, \textbf{5}, No.~2, 1304; \doi{10.1021/nn1030093}.

\bibitem{Peled2010}
Peled D., Naaman R., Daube S.S., J. Phys. Chem. B, 2010, \textbf{114}, No.~25,
  8581; \doi{10.1021/jp104533q}.

\bibitem{Heuer2013a}
Heuer-Jungemann A., Kirkwood R., El-Sagheer A.H., Brown T., Kanaras A.G.,
  Nanoscale, 2013, \textbf{5}, No.~16, 7209; \doi{10.1039/c3nr02362a}.

\bibitem{Chen2013}
Chen T., Wu C.S., Jimenez E., Zhu Z., Dajac J.G., You M., Han D., Zhang X., Tan
  W., Angew. Chem. Int. Edit., 2013, \textbf{52}, 2012; \doi{10.1002/anie.201209440}.

\bibitem{Zhang2012}
Zhang K., Hao L., Hurst S.J., Mirkin C.A., J. Am. Chem. Soc., 2012, \textbf{134},
  No.~40, 16488; \doi{10.1021/ja306854d}.

\bibitem{Seferos2007}
Seferos D.S., Giljohann D.A., Hill H.D., Prigodich A.E., Mirkin C.A., J. Am. Chem.
  Soc., 2007, \textbf{129}, No.~50, 15477; \doi{10.1021/ja0776529}.

\bibitem{Xin2009}
Xin A., Dong Q., Xiong C., Ling L., Chem. Commun., 2009,  No.~13, 1658; \doi{10.1039/b815825h}.

\bibitem{Lee2010}
Lee O.S., Prytkova T.R., Schatz G.C., J. Phys. Chem. Lett., 2010, \textbf{1},
  No.~12, 1781; \doi{10.1021/jz100435a}.

\bibitem{Zhang2013}
Zhang C., Macfarlane R.J., Young K.L., Choi C.H.J., Hao L., Auyeung E., Liu G.,
  Zhou X., Mirkin C.A., Nat. Mater., 2013, \textbf{12}, 741; \doi{10.1038/nmat3647}.

\bibitem{Zhang2013a}
Zhang Y., Lu F., Yager K.G., van~der Lelie D., Gang O., Nat. Nanotechnol.,
  2013, \textbf{8}, 865; \doi{10.1038/nnano.2013.209}.

\bibitem{vanderMeulen2013}
Van~der Meulen S.A., Leunissen M.E., J. Am. Chem. Soc., 2013, \textbf{135}, No.~40,
  15129; \doi{10.1021/ja406226b}.

\bibitem{Uberti2014}
Angioletti-Uberti S., Varilly P., Mognetti B.M., Frenkel~D., Phys. Rev. Lett.,
  2014, \textbf{113}, 128303; \doi{10.1103/PhysRevLett.113.128303}.

\bibitem{Lee2007}
Lee J.S.H., Lytton-Jean A.K., Hurst S.J., Mirkin C.A., Nano Lett., 2007,
  \textbf{7}, No.~7, 2112; \doi{10.1021/nl071108g}.

\bibitem{Han2012}
Han J., Ohara S., Sato K., Xu H., Tan Z., Morisada Y., Kuruma K., Naito M.,
  Shan P., Umetsu M., Mater. Lett., 2012, \textbf{79}, 78; \doi{10.1016/j.matlet.2012.03.094}.

\bibitem{Lee2006}
Lee J.S.H., Stoeva S.I., Mirkin C.A., J. Am. Chem. Soc., 2006, \textbf{128},
  No.~27, 8899; \doi{10.1021/ja061651j}.

\bibitem{Yang2006}
Yang J., Lee J.Y., Too H.P., Chow G.M., Gan L.M., Chem. Phys., 2006,
  \textbf{323}, No. 2-3, 304; \doi{10.1016/j.chemphys.2005.09.044}.

\bibitem{Chen2008}
Chen Y., Mao C., Small, 2008, \textbf{4}, No.~12, 2191; \doi{10.1002/smll.200800569}.

\bibitem{Kim2012}
Kim G.A., Han S.H., Lee J.S., Mater. Lett., 2012, \textbf{68}, 118; \doi{10.1016/j.matlet.2011.10.058}.

\bibitem{Kim2009a}
Kim J.Y., Lee J.S.H., Nano Lett., 2009, \textbf{9}, No.~12, 4564; \doi{10.1021/nl9030709}.

\bibitem{Casey2012}
Casey M.T., Scarlett R.T., Rogers W.B., Jenkins I., Sinno T., Crocker J.C., Nat.
  Commun., 2012, \textbf{3}, 1209; \doi{10.1038/ncomms2206}.

\bibitem{Nykypanchuk2008}
Nykypanchuk D., Maye M.M., van~der Lelie D., Gang O., Nature, 2008,
  \textbf{451}, No. 7178, 549; \doi{10.1038/nature06560}.

\bibitem{Dillenback2006}
Dillenback L.M., Goodrich G.P., Keating C.D., Nano Lett., 2006, \textbf{6},
  No.~1, 16; \doi{10.1021/nl0508873}.

\bibitem{Srivastava2013}
Srivastava S., Nykypanchuk D., Maye M.M., Tkachenko A.V., Gang O., Soft Matter,
  2013, \textbf{9}, No.~44, 10452; \doi{10.1039/c3sm51289d}.

\bibitem{Maye2010}
Maye M.M., Kumara M.T., Nykypanchuk D., Sherman W.B., Gang O., Nat.
  Nanotechnol., 2010, \textbf{5}, 116; \doi{10.1038/nnano.2009.378}.

\bibitem{Macfarlane2009}
Macfarlane R.J., Lee B., Hill H.D., Senesi A.J., Seifert S., Mirkin C.A., Proc.
  Natl. Acad. Sci. USA, 2009, \textbf{106}, No.~26, 10493; \doi{10.1073/pnas.0900630106}.

\bibitem{Senesi2013}
Senesi A.J., Eichelsdoerfer D.J., Macfarlane R.J., Jones M.R., Auyeung E., Lee
  B., Mirkin C.A., Angew. Chem. Int. Edit., 2013, \textbf{52}, No.~26, 6624; \doi{10.1002/anie.201301936}.

\bibitem{Auyeung2012}
Auyeung E., Cutler J.I., Macfarlane R.J., Jones M.R., Wu J., Liu G., Zhang K.,
  Osberg K.D., Mirkin C.A., Nat. Nanotechnol., 2012, \textbf{7}, 24; \doi{10.1038/nnano.2011.222}.

\bibitem{Keegan2013}
Keegan G.L., Aherne D., Defrancq E., Gun'ko Y.K., Kelly J.M.,
J. Phys. Chem. C, 2013, \textbf{117}, No.~1, 669; \doi{10.1021/jp309449d}.

\bibitem{Valignat2005}
Valignat M.P., Theodoly O., Crocker J.C., Russel W.B., Chaikin P.M., Proc. Natl.
  Acad. Sci. USA, 2005, \textbf{102}, No.~12, 4225; \doi{10.1073/pnas.0500507102}.

\bibitem{Shin2012}
Shin J., Zhang X., Liu J., J. Phys. Chem. B, 2012, \textbf{116}, No.~45, 13396; \doi{10.1021/jp310662m}.

\bibitem{Hill2008}
Hill H.D., Macfarlane R.J., Senesi A.J., Lee B., Park S.J., Mirkin C.A., Nano
  Lett., 2008, \textbf{8}, No.~8, 2341; \doi{10.1021/nl8011787}.

\bibitem{Storhoff2002}
Storhoff J.J., Elghanian R., Mirkin C.A., Letsinger R.L., Langmuir, 2002,
  \textbf{18}, No.~17, 6666; \doi{10.1021/la0202428}.

\bibitem{Doyen2013}
Doyen M., Bartik K., Bruylants G., Polymers, 2013, \textbf{5}, No.~3, 1041; \doi{10.3390/polym5031041}.

\bibitem{Park2008}
Park S.Y., Lytton-Jean A.K., Lee B., Weigand S., Schatz G.C., Mirkin C.A.,
  Nature, 2008, \textbf{451}, No. 7178, 553; \doi{10.1038/nature06508}.

\bibitem{Milam2003}
Milam V.T., Hiddessen A.L., Crocker J.C., Graves D.J., Hammer D.A., Langmuir,
  2003, \textbf{19}, No.~24, 10317; \doi{10.1021/la034376c}.

\bibitem{Lubitz2011}
Lubitz I., Kotlyar A., Bioconjugate Chem., 2011, \textbf{22}, No.~10, 2043; \doi{10.1021/bc200257e}.

\bibitem{Xiong2008}
Xiong H., van~der Lelie D., Gang O., J. Am. Chem. Soc., 2008, \textbf{130},
  No.~8, 2442; \doi{10.1021/ja710710j}.


\bibitem{Xiong2009}
Xiong H., van~der Lelie D., Gang O., Phys. Rev. Lett., 2009,
  \textbf{102}, 015504; \doi{10.1103/PhysRevLett.102.015504}.

\bibitem{Knorowski2011}
Knorowski C., Burleigh S., Travesset A., Phys. Rev. Lett., 2011,
  \textbf{106}, No.~21, 215501; \doi{10.1103/PhysRevLett.106.215501}.

\bibitem{Cigler2010}
Cigler P., Lytton-Jean A.K., Anderson D.G., Finn M.G., Park S.J., Nat.
  Mater., 2010, \textbf{9}, 918; \doi{10.1038/nmat2877}.

\bibitem{Park2001}
Park S.J., Lazarides A.A., Mirkin C.A., Letsinger R.L., Angew. Chem. Int. Edit.,
  2001, \textbf{40}, No.~15, 2909; \doi{10.1002/1521-3773(20010803)40:15<2909::AID-ANIE2909>3.0.CO;2-O}.

\bibitem{Park2004}
Park S.J., Lazarides A.A., Storhoff J.J., Pesce L., Mirkin C.A., J. Phys. Chem.~B,
  2004, \textbf{108}, No.~33, 12375; \doi{10.1021/jp040242b}.

\bibitem{Qin2005}
Qin W.J., Yung L.Y.L., Langmuir, 2005, \textbf{21}, No.~24, 11330; \doi{10.1021/la051630n}.

\bibitem{Strable2004}
Strable E., Johnson J.E., Finn M.G., Nano Lett., 2004, \textbf{4}, No.~8, 1385; \doi{10.1021/nl0493850}.

\bibitem{Sun2011}
Sun D., Gang O., J. Am. Chem. Soc., 2011, \textbf{133}, No.~14, 5252; \doi{10.1021/ja111542t}.

\bibitem{McGinley2013}
McGinley J.T., Jenkins I., Sinno T., Crocker J.C., Soft Matter, 2013,
  \textbf{9}, No.~38, 9119; \doi{10.1039/c3sm50950h}.

\bibitem{Nykypanchuk2007}
Nykypanchuk D., Maye M.M., van~der Lelie D., Gang O., Langmuir, 2007,
  \textbf{23}, No.~11, 6305; \doi{10.1021/la0637566}.

\bibitem{Kim2006}
Kim A.J., Biancaniello P.L., Crocker J.C., Langmuir, 2006, \textbf{22}, No.~5,
  1991; \doi{10.1021/la0528955}.

\bibitem{Kim2009}
Kim A.J., Scarlett R.T., Biancaniello P.L., Sinno T., Crocker J.C., Nat.
  Mater., 2009, \textbf{8}, 52; \doi{10.1038/nmat2338}.

\bibitem{Rogers2005}
Rogers P.H., Michel E., Bauer C.A., Vanderet S., Hansen D., Roberts B.K.,
  Calvez A., Crews J.B., Lau K.O., Wood A., Pine D.J., Schwartz P.V., Langmuir,
  2005, \textbf{21}, No.~12, 5562; \doi{10.1021/la046790y}.

\bibitem{Rogers2013}
Rogers W.B., Sinno T., Crocker J.C., Soft Matter, 2013, \textbf{9}, No.~28,
  6412; \doi{10.1039/c3sm50593f}.

\bibitem{Biancaniello2005}
Biancaniello P., Kim A., Crocker J., Phys. Rev. Lett., 2005,
  \textbf{94}, No.~5, 058302; \doi{10.1103/PhysRevLett.94.058302}.

\bibitem{Biancaniello2007}
Biancaniello P.L., Crocker J.C., Hammer D.A., Milam V.T., Langmuir, 2007,
  \textbf{23}, No.~5, 2688; \doi{10.1021/la062885j}.

\bibitem{Kegler2007}
Kegler K., Salomo M., Kremer F., Phys. Rev. Lett., 2007, \textbf{98},
  No.~5, 058304; \doi{10.1103/PhysRevLett.98.058304}.

\bibitem{Kegler2008}
Kegler K., Konieczny M., Dominguez-Espinosa G., Gutsche C., Salomo M., Kremer
  F., Likos C.N., Phys. Rev. Lett., 2008, \textbf{100}, No.~11, 118302; \doi{10.1103/PhysRevLett.100.118302}.

\bibitem{Chollakup2010}
Chollakup R., Smitthipong W., Chworos A., Polym. Chem., 2010, \textbf{1},
  No.~5, 658; \doi{10.1039/b9py00290a}.

\bibitem{Scarlett2010}
Scarlett R.T., Crocker J.C., Sinno T., J. Chem. Phys., 2010, \textbf{132}, No.~23,
  234705; \doi{10.1063/1.3453704}.

\bibitem{Scarlett2011}
Scarlett R.T., Ung M.T., Crocker J.C., Sinno T., Soft Matter, 2011, \textbf{7},
  No.~5, 1912; \doi{10.1039/c0sm00370k}.

\bibitem{Leunissen2009}
Leunissen M.E., Dreyfus R., Cheong F.C., Grier D.G., Sha R., Seeman N.C.,
  Chaikin P.M., Nat. Mater., 2009, \textbf{8}, No.~7, 590; \doi{10.1038/nmat2471}.

\bibitem{Feng2012}
Feng L., Sha R., Seeman N.C., Chaikin P.M., Phys. Rev. Lett., 2012,
  \textbf{109}, No.~18, 188301; \doi{10.1103/PhysRevLett.109.188301}.

\bibitem{Claridge2005}
Claridge S.A., Goh S.L., Frechet J.M.J., Williams S.C., Micheel C.M.,
  Alivisatos A.P., Chem. Mater., 2005, \textbf{17}, No.~7, 1628; \doi{10.1021/cm0484089}.

\bibitem{Maye2006}
Maye M.M., Nykypanchuk D., van~der Lelie D., Gang O., J. Am. Chem. Soc., 2006,
  \textbf{128}, No.~43, 14020; \doi{10.1021/ja0654229}.

\bibitem{Maye2007}
Maye M.M., Nykypanchuk D., van~der Lelie D., Gang O., Small, 2007, \textbf{3},
  No.~10, 1678; \doi{10.1002/smll.200700357}.

\bibitem{Cutler2011}
Cutler J.I., Zhang K., Zheng D., Auyeung E., Prigodich A.E., Mirkin C.A., J. Am.
  Chem. Soc., 2011, \textbf{133}, No.~24, 9254; \doi{10.1021/ja203375n}.

\bibitem{Jones2011}
Jones M.R., Macfarlane R.J., Prigodich A.E., Patel P.C., Mirkin C.A., J. Am. Chem.
  Soc., 2011, \textbf{133}, No.~46, 18865; \doi{10.1021/ja206777k}.

\bibitem{Jones2010}
Jones M.R., Macfarlane R.J., Lee B., Zhang J., Young K.L., Senesi A.J., Mirkin
  C.A., Nat. Mater., 2010, \textbf{9}, 913; \doi{10.1038/nmat2870}.

\bibitem{Dujardin2001}
Dujardin E., Hsin L.B., Wang C., Mann S., Chem. Commun., 2001, No.~14, 1264; \doi{10.1039/b102319p}.

\bibitem{Geerts2013}
Geerts N., Schreck C.F., Beales P.A., Shigematsu H., O'Hern C.S., Vanderlick
  T.K., Langmuir, 2013, \textbf{29}, No.~42, 13089; \doi{10.1021/la403091w}.

\bibitem{Tan2013}
Tan L.H., Xing H., Chen H., Lu Y., J. Am. Chem. Soc., 2013, \textbf{135}, No.~47,
  17675; \doi{10.1021/ja408033e}.

\bibitem{Qin2008}
Qin W.J., Yung L.Y.L., Bioconjugate Chem., 2008, \textbf{19}, 385; \doi{10.1021/bc700178f}.

\bibitem{Zhang2006}
Zhang J., Liu Y., Ke Y., Yan H., Nano Lett., 2006, \textbf{6}, No.~2, 248; \doi{10.1021/nl052210l}.

\bibitem{Pinto2005}
Pinto Y.Y., Le J.D., Seeman N.C., Musier-Forsyth K., Taton T.A., Kiehl R.A.,
  Nano Lett., 2005, \textbf{5}, No.~12, 2399; \doi{10.1021/nl0515495}.

\bibitem{Millstone2008}
Millstone J.E., Georganopoulou D.G., Xu X.Y., Wei W., Li S.Y., Mirkin C.A.,
  Small, 2008, \textbf{4}, No.~12, 2176; \doi{10.1002/smll.200800931}.

\bibitem{Young2012}
Young K.L., Jones M.R., Zhang J., Macfarlane R.J., Esquivel-Sirvent R., Nap
  R.J., Wu J., Schatz G.C., Lee B., Mirkin C.A., Proc. Natl. Acad. Sci. USA,
  2012, \textbf{109}, No.~7, 2240; \doi{10.1073/pnas.1119301109}.

\bibitem{Maye2009}
Maye M.M., Nykypanchuk D., Cuisinier M., van~der Lelie D., Gang O., Nat.
  Mater., 2009, \textbf{8}, 388; \doi{10.1038/nmat2421}.

\bibitem{Xing2012}
Xing H., Wang Z., Xu Z., Xiang Y., Liu G.L., Lu Y., ACS Nano, 2012, \textbf{6},
  No.~1, 802; \doi{10.1021/nn2042797}.

\bibitem{Xu2006}
Xu X., Rosi N.L., Wang Y., Huo F., Mirkin C.A., J. Am. Chem. Soc., 2006,
  \textbf{128}, No.~29, 9286; \doi{10.1021/ja061980b}.

\bibitem{Cederquist2009}
Cederquist K.B., Keating C.D., ACS Nano, 2009, \textbf{3}, No.~2, 256; \doi{10.1021/nn9000726}.

\bibitem{Vial2013}
Vial S., Nykypanchuk D., Yager K.G., Tkachenko A.V., Gang O., ACS Nano, 2013,
  \textbf{7}, No.~6, 5437; \doi{10.1021/nn401413b}.

\bibitem{Jahn2010}
Jahn S., Geerts N., Eiser E., Langmuir, 2010, \textbf{26}, No.~22, 16921; \doi{10.1021/la103192q}.

\bibitem{Cheng2009}
Cheng W., Campolongo M.J., Cha J.J., Tan S.J., Umbach C.C., Muller D.A., Luo
  D., Nat. Mater., 2009, \textbf{8}, No.~8, 519; \doi{10.1038/nmat2440}.

\bibitem{Lan2013}
Lan X., Chen Z., Liu B.J., Ren B., Henzie J., Wang Q., Small, 2013, \textbf{9},
  No.~13, 2308; \doi{10.1002/smll.201202503}.

\bibitem{Geerts2010a}
Geerts N., Eiser E., Soft Matter, 2010, \textbf{6}, No.~3, 664; \doi{10.1039/B917846E}.

\bibitem{Estephan2013}
Estephan Z.G., Qian Z., Lee D., Crocker J.C., Park S.J., Nano Lett., 2013,
  \textbf{13}, No.~9, 4449; \doi{10.1021/nl4023308}.

\bibitem{Hung2010a}
Hung A.M., Micheel C.M., Bozano L.D., W. O.L., Wallraff G.M., Cha J.N., Nat.
  Nanotechnol., 2010, \textbf{5}, 121; \doi{10.1038/nnano.2009.450}.

\bibitem{Taton2000}
Taton T.A., Mucic R.C., Mirkin C.A., Letsinger R.L., J. Am. Chem. Soc., 2000,
  \textbf{122}, No.~26, 6305; \doi{10.1021/ja0007962}.

\bibitem{Shyr2008}
Shyr M.H.S., Wernette D.P., Wiltzius P., Lu Y., Braun P.V., J. Am. Chem. Soc.,
  2008, \textbf{130}, No.~26, 8234; \doi{10.1021/ja711026r}.

\bibitem{DeMille2011}
DeMille R.C., Cheatham~III T.E., Molinero V., J. Phys. Chem. B, 2011,
  \textbf{115}, No.~1, 132; \doi{10.1021/jp107028n}.

\bibitem{Kenward2009}
Kenward M., Dorfman K.D., J. Chem. Phys., 2009, \textbf{130}, No.~9, 095101; \doi{10.1063/1.3078795}.

\bibitem{Drukker2001}
Drukker K., Wu G., Schatz G.C., J. Chem. Phys., 2001,
  \textbf{114}, No.~1, 579; \doi{10.1063/1.1329137}.

\bibitem{Knotts2007}
Knotts T.A.t., Rathore N., Schwartz D.C., de~Pablo J.J., J. Chem. Phys., 2007,
  \textbf{126}, No.~8, 084901; \doi{10.1063/1.2431804}.

\bibitem{Sambriski2009}
Sambriski E.J., Schwartz D.C., de~Pablo J.J., Biophys. J., 2009, \textbf{96},
  No.~5, 1675; \doi{10.1016/j.bpj.2008.09.061}.

\bibitem{Lequieu2015}
Lequieu J.P., Hinckley D.M., de~Pablo J.J., Soft Matter, 2015, \textbf{11}, 1919; \doi{10.1039/C4SM02573C}.

\bibitem{Linak2011}
Linak M.C., Tourdot R., Dorfman K.D., J. Chem. Phys., 2011, \textbf{135}, No.~20,
  205102; \doi{10.1063/1.3662137}.

\bibitem{Mladek2013}
Mladek B.M., Fornleitner J., Martinez-Veracoechea F.J., Dawid A., Frenkel~D.,
  Soft Matter, 2013, \textbf{9}, No.~30, 7342; \doi{10.1039/c3sm50701g}.

\bibitem{Ding2014}
Ding Y., Mittal J., J. Chem. Phys., 2014, \textbf{141}, No.~18,
  184901; \doi{10.1063/1.4900891}.

\bibitem{Sales2005}
Sales-Pardo M., Guimer\'a R., Moreira A., Widom J., Amaral L., Phys. Rev.
  E, 2005, \textbf{71}, No.~5, 051902; \doi{10.1103/PhysRevE.71.051902}.

\bibitem{Starr2006}
Starr F.W., Sciortino F., J. Phys.: Condens. Matter, 2006, \textbf{18}, No.~26,
  L347; \doi{10.1088/0953-8984/18/26/L02}.

\bibitem{Theodorakis2013}
Theodorakis P.E., Dellago C., Kahl G., J. Chem. Phys., 2013, \textbf{138}, No.~2,
  025101; \doi{10.1063/1.4773920}.

\bibitem{Vargas2011}
Lara V.F., Starr F.W., Soft Matter, 2011, \textbf{7}, No.~5, 2085; \doi{10.1039/c0sm00989j}.

\bibitem{Padovan2011}
Padovan-Merhar O., Lara F.V., Starr F.W., J. Chem. Phys., 2011, \textbf{134},
  No.~24, 244701; \doi{10.1063/1.3596745}.

\bibitem{Largo2007}
Largo J., Starr F.W., Sciortino F., Langmuir, 2007, \textbf{23}, No.~11, 5896; \doi{10.1021/la063036z}.

\bibitem{Dai2010}
Dai W., Hsu C.W., Sciortino F., Starr F.W., Langmuir, 2010, \textbf{26}, No.~5,
  3601; \doi{10.1021/la903031p}.

\bibitem{Dai2010a}
Dai W., Kumar S.K., Starr F.W., Soft Matter, 2010, \textbf{6}, No.~24, 6130; \doi{10.1039/c0sm00484g}.

\bibitem{Hsu2008}
Hsu C.W., Largo J., Sciortino F., Starr F.W., Proc. Natl. Acad. Sci. USA, 2008,
  \textbf{105}, No.~37, 13711; \doi{10.1073/pnas.0804854105}.

\bibitem{Flory1953}
Flory P., Principles of Polymer Chemistry, Cornell University Press, New York,
  1953.

\bibitem{Stockmayer1943}
Stockmayer W.H., J. Chem. Phys., 1943, \textbf{11}, No.~45; \doi{10.1063/1.1723803}.

\bibitem{Hsu2010}
Hsu C.W., Sciortino F., Starr F.W., Phys. Rev. Lett., 2010,
  \textbf{105}, No.~5, 055502; \doi{10.1103/PhysRevLett.105.055502}.

\bibitem{Seifpour2013}
Seifpour A., Dahl S.R., Jayaraman A., Mol. Simulat.,  2014, \textbf{40}, No.~14, 1085; \doi{10.1080/08927022.2013.845888}.

\bibitem{Seifpour2013a}
Seifpour A., Dahl S.R., Lin B., Jayaraman A., Mol. Simulat., 2013,
  \textbf{39}, No.~9, 741; \doi{10.1080/08927022.2013.765569}.

\bibitem{Chi2012}
Chi C., Lara V.F., Tkachenko A.V., Starr F.W., Gang O., ACS Nano, 2012,
  \textbf{6}, No.~8, 6793; \doi{10.1021/nn301528h}.

\bibitem{Rovigatti2014}
Rovigatti L., Smallenburg F., Romano F., Sciortino F., ACS Nano, 2014,
  \textbf{8}, No.~4, 3567; \doi{10.1021/nn501138w}.

\bibitem{Rovigatti2014b}
Rovigatti L., Bomboi F., Sciortino F., J. Chem. Phys., 2014, \textbf{140},
  154903; \doi{10.1063/1.4870467}.

\bibitem{Knorowski2011a}
Knorowski C., Travesset A., Curr. Opin. Solid State Mat. Sci., 2011, \textbf{15}, No.~6, 262; \doi{10.1016/j.cossms.2011.07.002}.

\bibitem{Knorowski2012}
Knorowski C., Travesset A., Soft Matter, 2012, \textbf{8}, No.~48, 12053; \doi{10.1039/c2sm26832a}.

\bibitem{Li2012}
Li T.I., Sknepnek R., Macfarlane R.J., Mirkin C.A., de~la Cruz~M.O., Nano Lett.,
  2012, \textbf{12}, No.~5, 2509; \doi{10.1021/nl300679e}.

\bibitem{Li2013}
Li T.I., Sknepnek R., de~la Cruz M.O., J. Am. Chem. Soc., 2013, \textbf{135},
  No.~23, 8535; \doi{10.1021/ja312644h}.

\bibitem{Auyeung2014}
Auyeung E., Li T., Senesi A.J., Schmucker A.L., Pals B.C., de~la Cruz~M.O., Mirkin C.A., Nature, 2014, \textbf{505}, 73; \doi{10.1038/nature12739}.

\bibitem{Leunissen2011}
Leunissen M.E., Frenkel~D., J. Chem. Phys., 2011, \textbf{134}, No.~8, 084702; \doi{10.1063/1.3557794}.

\bibitem{Guerrini2013}
Guerrini L., Barrett L., Dougan J.A., Faulds K., Graham D., Nanoscale, 2013,
  \textbf{5}, No.~10, 4166; \doi{10.1039/c3nr011}.

\bibitem{Mognetti2012}
Mognetti B.M., Leunissen M.E., Frenkel~D., Soft Matter, 2012, \textbf{8},
  No.~7, 2213; \doi{10.1039/c2sm06635a}.

\bibitem{Uberti2012}
Uberti-Angioletti S., Mognetti B.M., Frenkel~D., Nat. Mater., 2012,
  \textbf{11}, 518; \doi{10.1038/nmat3314}.

\bibitem{Gang2012}
Gang O., Nat. Mater., 2012, \textbf{11}, No.~6, 487; \doi{10.1038/nmat3344}.

\bibitem{DiMichele2013a}
Di~Michele L., Varrato F., Kotar J., Nathan S.H., Foffi G., Eiser E., Nat.
  Commun., 2013, \textbf{4}, 2007; \doi{10.1038/ncomms3007}.

\bibitem{Martinez2010}
Martinez-Veracoechea F.J., Bozorgui B., Frenkel~D., Soft Matter, 2010,
  \textbf{6}, No.~24, 6136; \doi{10.1039/c0sm00567c}.

\bibitem{Martinez2011}
Martinez-Veracoechea F.J., Mladek B.M., Tkachenko A.V., Frenkel~D., Phys.
  Rev. Lett., 2011, \textbf{107}, No.~4, 045902; \doi{10.1103/PhysRevLett.107.045902}.

\bibitem{Hazarika2007}
Hazarika P., Irrgang J., Spengler M., Niemeyer C.M., Adv. Funct.
  Mater., 2007, \textbf{17}, No.~3, 437; \doi{10.1002/adfm.200600694}.

\bibitem{Mladek2012}
Mladek B.M., Fornleitner J., Martinez-Veracoechea F.J., Dawid A., Frenkel~D.,
  Phys. Rev. Lett., 2012, \textbf{108}, No.~26, 268301; \doi{10.1103/PhysRevLett.108.268301}.

\bibitem{Rabideau2007}
Rabideau B.D., Bonnecaze R.T., Langmuir, 2007, \textbf{23}, No.~20, 10000; \doi{10.1021/la701166p}.

\bibitem{Yanagishima2014}
Yanagishima T., Laohakunakorn N., Keyser U.F., Eiser E., Tanaka H., Soft
  Matter, 2014, \textbf{10}, No.~11, 1738; \doi{10.1039/c3sm52830h}.

\bibitem{Schmatko2007}
Schmatko T., Bozorgui B., Geerts N., Frenkel~D., Eiser E., Poon W.C.K., Soft
  Matter, 2007, \textbf{3}, No.~6, 703; \doi{10.1039/b618028k}.

\bibitem{Dreyfus2010}
Dreyfus R., Leunissen M.E., Sha R., Tkachenko A., Seeman N.C., Pine D.J.,
  Chaikin P.M., Phys. Rev. E, 2010, \textbf{81}, No.~4, 041404; \doi{10.1103/PhysRevE.81.041404}.

\bibitem{Rogers2011}
Rogers B.W., Crocker J.C., Proc. Natl. Acad. Sci. USA, 2011, \textbf{108},
  No.~38, 15687; \doi{10.1073/pnas.1109853108}.

\bibitem{Mognetti2012a}
Mognetti B.M., Varilly P., Angioletti-Uberti S., Martinez-Veracoechea F.J.,
  Dobnikar J., Leunissen M.E., Frenkel~D., Proc. Natl. Acad. Sci. USA, 2012,
  \textbf{109}, No.~7, E378; \doi{10.1073/pnas.1119991109}.

\bibitem{Dreyfus2009}
Dreyfus R., Leunissen M., Sha R., Tkachenko A., Seeman N., Pine D., Chaikin P.,
  Phys. Rev. Lett., 2009, \textbf{102}, No.~4, 048301; \doi{10.1103/PhysRevLett.102.048301}.

\bibitem{Sun2005}
Sun Y., Harris N.C., Kiang C.H., Physica A, 2005, \textbf{350}, No.~1, 89; \doi{10.1016/j.physa.2005.01.013}.

\bibitem{Park2003a}
Park S., Stroud D., Phys. Rev. B, 2003, \textbf{68}, 224201; \doi{10.1103/PhysRevB.68.224201}.

\bibitem{Park2003}
Park S., Stroud D., Phys. Rev. B, 2003, \textbf{67}, 212202; \doi{10.1103/PhysRevB.67.212202}.

\bibitem{Lazarides2000}
Lazarides A.A., Schatz G.C., J. Chem. Phys., 2000,
  \textbf{112}, No.~6, 2987; \doi{10.1063/1.480873}.

\bibitem{Lazarides2000a}
Lazarides A.A., Schatz G.C., J. Phys. Chem. B, 2000, \textbf{104}, No.~3, 460; \doi{10.1021/jp992179+}.

\bibitem{Yan2012}
Yan Y., Chen J.I., Ginger D.S., Nano Lett., 2012, \textbf{12}, No.~5, 2530; \doi{10.1021/nl300739n}.

\bibitem{Severac2012}
S\'everac F., Alphonse P., Est\`eve A., Bancaud A., Rossi C., Adv.
  Funct. Mater., 2012, \textbf{22}, No.~2, 323; \doi{10.1002/adfm.201100763}.

\bibitem{Busson2011}
Busson M.P., Rolly B., Stout B., Bonod N., Larquet E., Polman A., Bidault S.,
  Nano Lett., 2011, \textbf{11}, No.~11, 5060; \doi{10.1021/nl2032052}.

\bibitem{Mukundan2013}
Mukundan V.T., Quang M.N.T., Miao Y.H., Phan A.T., Soft Matter, 2013,
  \textbf{9}, No.~1, 216; \doi{10.1039/C2SM26652K}.

\bibitem{Varilly2012}
Varilly P., Angioletti-Uberti S., Mognetti B.M., Frenkel~D., J. Chem. Phys., 2012,
  \textbf{137}, No.~9, 094108; \doi{10.1063/1.4748100}.

\bibitem{Angioletti-Uberti2013}
Angioletti-Uberti~S., Varilly~P., Mognetti~B.M., Tkachenko~A.V., Frenkel~D., J. Chem. Phys., 2013,  \textbf{138},
021102; \doi{10.1063/1.4775806}.


\bibitem{Tkachenko2002}
Tkachenko A.V., Phys. Rev. Lett., 2002, \textbf{89}, No.~14, 148303; \doi{10.1103/PhysRevLett.89.148303}.

\bibitem{Tkachenko2011}
Tkachenko A.V., Phys. Rev. Lett., 2011, \textbf{106}, 255501; \doi{10.1103/PhysRevLett.106.255501}.

\bibitem{Halverson2013}
Halverson J.D., Tkachenko A.V., Phys. Rev. E, 2013, \textbf{87}, No.~6, 062310; \doi{10.1103/PhysRevE.87.062310}.

\bibitem{Zeravcic2014}
Zeravcic Z., Manoharan V.N., Brenner M.P., Proc. Natl. Acad. Sci. USA, 2014,
  \doi{10.1073/pnas.1411765111}.

\bibitem{Licata2006}
Licata N., Tkachenko A., Phys. Rev. E, 2006, \textbf{74}, No.~4., 040401(R); \doi{10.1103/PhysRevE.74.040401}.

\bibitem{Licata2006a}
Licata N., Tkachenko A., Phys. Rev. E, 2006, \textbf{74}, No.~4., 041406; \doi{10.1103/PhysRevE.74.041406}.

\bibitem{Licata2008}
Licata N.A., Tkachenko A.V., Europhys. Lett., 2008, \textbf{81},
  No.~4, 48009; \doi{10.1209/0295-5075/81/48009}.

\bibitem{Licata2009}
Licata N., Tkachenko A., Phys. Rev. E, 2009, \textbf{79}, No.~1, 011404; \doi{10.1103/PhysRevE.79.011404}.

\bibitem{Jenkins2014}
Jenkins I.C., Casey M.T., McGinley J.T., Crocker J.C., Sinno T., Proc. Natl. Acad.
  Sci. USA, 2014, \textbf{111}, No.~13, 4803; \doi{10.1073/pnas.1318012111}.

\bibitem{Lukatsky2004}
Lukatsky D., Frenkel~D., Phys. Rev. Lett., 2004, \textbf{92}, No.~6, 068302; \doi{10.1103/PhysRevLett.92.068302}.

\bibitem{Lukatsky2005}
Lukatsky D.B., Frenkel~D., J. Chem. Phys., 2005, \textbf{122}, No.~21, 214904; \doi{10.1063/1.1906210}.

\bibitem{Lukatsky2006}
Lukatsky D.B., Mulder B.M., Frenkel~D., J. Phys.: Condens. Matter,
  2006, \textbf{18}, No.~18, S567; \doi{10.1088/0953-8984/18/18/S05}.

\bibitem{Jin2003a}
Jin R., Wu G., Li Z., Mirkin C.A., Schatz G.C., J. Am. Chem. Soc., 2013,
  \textbf{125}, No.~6, 1643; \doi{10.1021/ja021096v}.

\bibitem{Talanquer2006}
Talanquer V., J. Chem. Phys., 2006, \textbf{125}, No.~19, 194701; \doi{10.1063/1.2370872}.

\bibitem{Tindemans2010}
Tindemans S.H., Mulder B.M., Phys. Rev. E, 2010, \textbf{82}, No.~2, 021404; \doi{10.1103/PhysRevE.82.021404}.

\bibitem{Srinivasan2013}
Srinivasan B., Vo T., Zhang Y., Gang O., Kumar S., Venkatasubramanian V., Proc.
  Natl. Acad. Sci. USA, 2013, \textbf{110}, No.~46, 18431; \doi{10.1073/pnas.1316533110}.

\bibitem{Gottwald2005}
Gottwald D., Kahl G., Likos C.N., J. Chem. Phys., 2005, \textbf{122}, 204503; \doi{10.1063/1.1901585}.

\bibitem{Fornleitner2008}
Fornleitner J., Lo~Verso F., Kahl G., Likos C.N., Soft Matter, 2008,
  \textbf{4}, 480; \doi{10.1039/b717205b}.

\bibitem{Doppelbauer2010}
Doppelbauer G., Bianchi E., Kahl G., J. Phys.: Condens. Matter, 2010, \textbf{22},
  104105; \doi{10.1088/0953-8984/22/10/104105}.

\bibitem{Doppelbauer2012}
Doppelbauer G., Noya E.G., Bianchi E., Kahl G., Soft Matter, 2012, \textbf{8},
  7768; \doi{10.1039/c2sm26043c}.

\bibitem{Lee2009}
Lee O.S., Schatz G.C., J. Phys. Chem. C, 2009, \textbf{113}, No.~6, 2316; \doi{10.1021/jp8094165}.

\bibitem{Ngo2012}
Ngo V.A., Kalia R.K., Nakano A., Vashishta P., J. Phys.
  Chem. C, 2012, \textbf{116}, No.~36, 19579; \doi{10.1021/jp306133v}.

\bibitem{Knorowski2014}
Knorowski C., Travesset A., J. Am. Chem. Soc., 2014, \textbf{136}, No.~2, 653; \doi{10.1021/ja406241n}.

\bibitem{Dhakal2013}
Dhakal S., Kohlstedt K.L., Schatz G.C., Mirkin C.A., de~la Cruz~M.O., ACS Nano,
  2013, \textbf{7}, No.~12, 10948; \doi{10.1021/nn404476f}.

\bibitem{Leunissen2010}
Leunissen M.E., Dreyfus R., Sha R., Seeman N.C., Chaikin P.M., J. Am. Chem. Soc.,
  2010, \textbf{132}, No.~6, 1903; \doi{10.1021/ja907919j}.

\bibitem{Noyong2006}
Noyong M., Ceyhan B., Niemeyer C.M., Simon U., Colloid Polym. Sci., 2006,
  \textbf{284}, 1265; \doi{10.1007/s00396-006-1518-3}.

\bibitem{Cobbe2003}
Cobbe S., Connolly S., Ryan D., Nagle L., Eritja R., Fitzmaurice D., J. Phys.
  Chem. B, 2003, \textbf{107}, No.~2, 470; \doi{10.1021/jp021503p}.

\bibitem{Diaz2012}
Diaz J.A., Grewer D.M., Gibbs-Davis J.M., Small, 2012, \textbf{8}, No.~6, 873; \doi{10.1002/smll.201101922}.

\bibitem{Jayagopal2010}
Jayagopal A., Halfpenny K.C., Perez J.W., Wright D.W., J. Am. Chem. Soc., 2010,
  \textbf{132}, No.~28, 9789; \doi{10.1021/ja102585v}.

\bibitem{Leunissen2009a}
Leunissen M.E., Dreyfus R., Sha R., Wang T., Seeman N.C., Pine D.J., Chaikin
  P.M., Soft Matter, 2009, \textbf{5}, No.~12, 2422; \doi{10.1039/b817679e}.

\bibitem{Li2010}
Li D., Banon S., Biswal S.L., Soft Matter, 2010, \textbf{6}, No.~17, 4197; \doi{10.1039/c0sm00159g}.

\bibitem{Park2006}
Park S.J., Lee J.S.H., Georganopoulou D., Mirkin C.A., Schatz G.C., J. Phys. Chem.
  B, 2006, \textbf{110}, No.~25, 12673; \doi{10.1021/jp062212+}.

\bibitem{Wu2013}
Wu K.T., Feng L., Sha R., Dreyfus R., Grosberg A.Y., Seeman N.C., Chaikin P.M.,
  Phys. Rev. E, 2013, \textbf{88}, No.~2, 022304; \doi{10.1103/PhysRevE.88.022304}.

\bibitem{Auyeung2012b}
Auyeung E., Macfarlane R.J., Choi C.H.J., Cutler J.I., Mirkin C.A., Adv.
  Mater., 2012, \textbf{24}, No.~38, 5181; \doi{10.1002/adma.201202069}.

\bibitem{Radha2014}
Radha B., Senesi A.J., O'Brien M.N., Wang M.X., Auyeung E., Lee B., Mirkin
  C.A., Nano Lett., 2014, \textbf{14}, No.~4, 2162; \doi{10.1021/nl500473t}.

\bibitem{Kim2013}
Kim Y., Macfarlane R.J., Mirkin C.A., J. Am. Chem. Soc., 2013, \textbf{135},
  No.~28, 10342; \doi{10.1021/ja405988r}.

\end{thebibliography}
\end{document}